\documentclass[11pt,numbers]{article}
\usepackage[square,comma,sort&compress]{natbib}
\usepackage{graphicx}
\usepackage{amssymb}
\usepackage{fixltx2e}
\usepackage{amsmath}
\usepackage{amstext}
\usepackage[]{units}
\usepackage{ulem}
\usepackage[T1]{fontenc}
\usepackage[utf8]{inputenc}
\usepackage{authblk}

\begin{document}


\title{Wave mitigation in ordered networks of granular chains}

\author[1]{Andrea Leonard}
\author[2]{Laurent Ponson}
\author[3,4]{Chiara Daraio}

\affil[1]{Department of Mechanical and Civil Engineering, Pasadenda, CA 91125, USA}
\affil[2]{Institut Jean le Rond d'Alembert (UMR 7190), CNRS - Universit\'{e}, Pierre et Marie Curie, 75005 Paris, France}
\affil[3]{Graduate Aerospace Laboratories (GALCIT), Pasadenda, CA 91125, USA}
\affil[4]{Department of Mechanical and Process Engineering (D-MAVT), ETH--Zurich, Zurich, Switzerland 8092}

\renewcommand\Authands{ and }
 \date{}

\maketitle

\begin{abstract}
We study the propagation of stress waves through ordered 2D networks of granular chains. The quasi-particle continuum theory employed captures the acoustic pulse splitting, bending, and recombination through the network and is used to derive its effective acoustic properties. The strong wave mitigation properties of the network predicted theoretically are confirmed through both numerical simulations and experimental tests. In particular, the leading pulse amplitude propagating through the system is shown to decay exponentially with the propagation distance and the spatial structure of the transmitted wave shows an exponential localization along the direction of the incident wave. The length scales that characterized these exponential decays are studied and determined as a function of the geometrical properties of the network. These results open avenues for the design of efficient impact mitigating structures and provide new insights into the mechanisms of wave propagation in granular matter.
\end{abstract}



\section{Introduction}
\label{Introduction}
The design of metamaterials with unusual effective properties with respect to natural materials has driven a large amount of research for about a decade. Metamaterials, defined as materials with engineered properties, were first conceived to achieve specific macroscopic electromagnetic and optical behaviors not readily occurring in nature \cite{Pendry,Shelby}. Driven by the mathematical analogy between acoustic and electromagnetic waves, researchers have been looking more recently for acoustic counterparts of these systems: linear acoustic metamaterials with negative effective Young's modulus or mass density have produced interesting acoustic behaviors, showing many promises for engineering applications \cite{Fang,Brun,Huang}. In particular, some of these systems can prevent the propagation of harmonic waves, behaving like super absorbers with a transmission coefficient close to zero in some range of frequencies \cite{Huang2,Mei}. Even though such band gaps can be chosen adequately by tuning the metamaterial microstructure \cite{Popa}, wave mitigation is usually only efficient in a limited range of frequencies. As a result, these systems are relevant for some specific types of incident acoustic waves and might be inappropriate for absorbing impacts or complex excitations that generate a broad spectrum of harmonic waves.

\begin{figure}[h]
\includegraphics[scale=0.65]{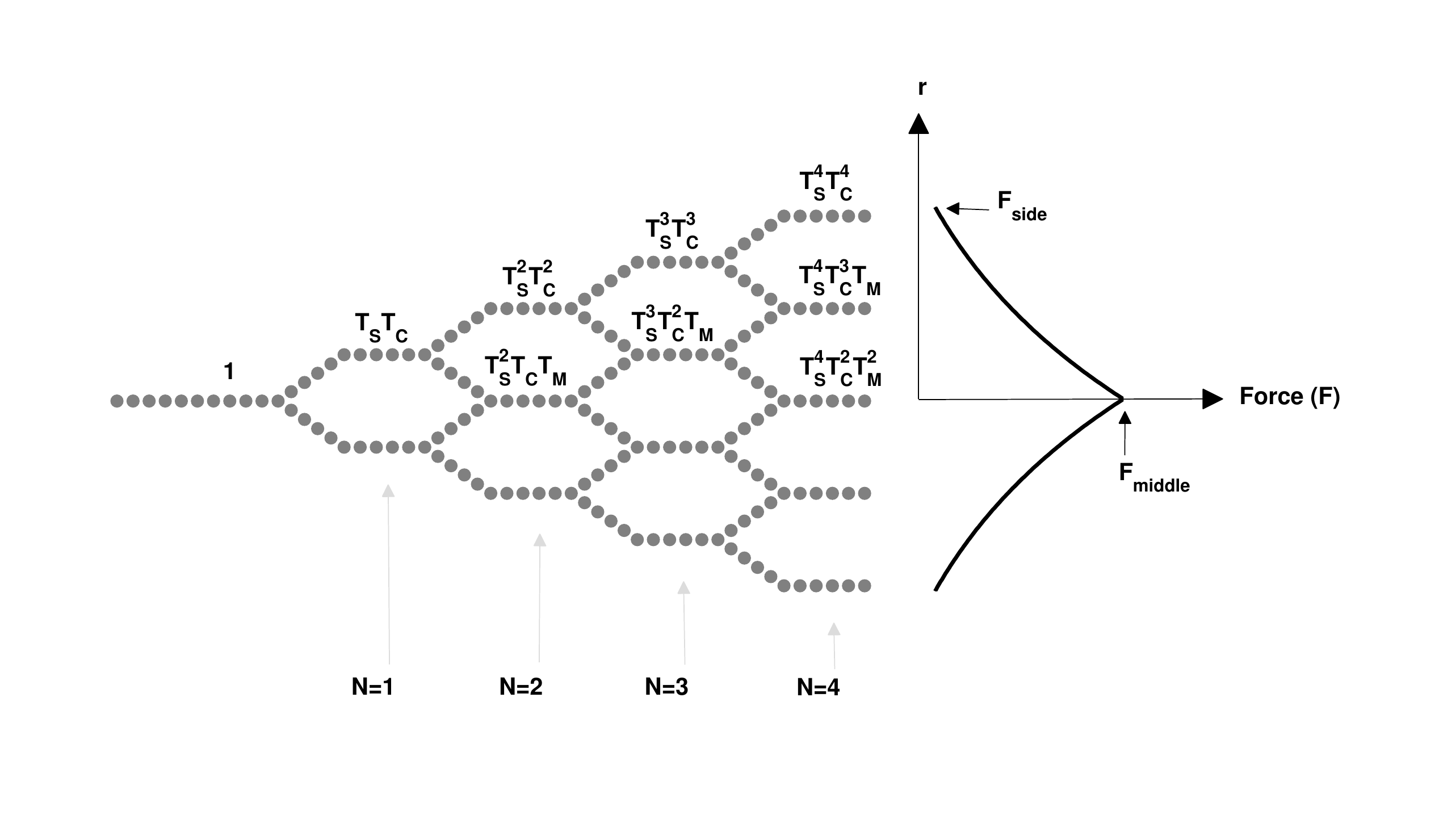}
\centering
\caption{Schematic diagram of the granular network investigated in this study. (Left) The wave transmission coefficients at each of the straight segments is given as a function of elementary transmission coefficients , $T_{\mathrm S}$, $T_{\mathrm C}$ and $T_{\mathrm M}$, and corresponds to N=1 up to N=4 levels structures. (Right) The transmitted force $F$ profile as a function of the normalized distance $r$ from the middle of the network.}\label{description}
\end{figure}

To overcome this difficulty, we use here a nonlinear acoustic medium as a building brick for the design of an efficient absorber. In the networks of granular chains like the one shown on Fig. \ref{description}, the acoustic energy can propagate only through highly nonlinear waves with prescribed properties imposed by the bead properties \cite{Nesterenko:1983,FNesterenko:2001} \--- the only variable parameter of these solitary waves is their total acoustic energy that governs their force amplitude, their traveling velocity, etc... This allows us to control and manipulate the transport of mechanical energy in the system. Following this idea, we have designed an acoustic metamaterial supporting only non-linear solitary waves that produces a robust acoustic response, independent of the nature of the incident wave and capable of efficiently mitigating complex dynamic excitations.

In addition, we take advantage of the network structure of granular materials to guide acoustic waves along predefined paths, and split, recombine and redirect waves as desired using an engineered network of granular chains without dissipating energy within the shielding system. In particular, we show that wave splitting is extremely efficient for mitigating acoustic waves, without calling for the dissipative processes usually present in most of acoustic media \--- the effect of which will actually sum up with the mitigation produced by wave splitting. Additionally, we compare the performance of the granular networks with that of a non-dispersive continuum media, which clearly demonstrates the system efficiency in mitigating.

Here, both the nonlinear response and the underlying branched structure of the granular networks are used to produce highly stress mitigating and robust metamaterials. The dynamical behavior of the designed network is investigated using a hybrid experimental and theoretical approach. In particular, we use the theoretical framework describing solitary waves pioneered by \cite{Nesterenko:1983} to predict quantitatively wave propagation through the metamaterial, and bridge the network geometry with its effective acoustic properties. Extensive studies \cite{Coste:1997,Daraio:2005,Sen:2008} have shown that uncompressed chains of spheres support traveling solitary waves, the behavior of which can be elegantly predicted using the so-called quasi-particle model based on energy and momentum conservation \cite{FNesterenko:2001,Job:2007,Daraio:2010}. Here, we use this approach to derive the transmission coefficients required to describe wave propagation through the bends and junctions of the granular network. These findings serve as elementary bricks to predict the evolution of the main pulse amplitude traveling through the network as well as the spatial repartition of transmitted waves in the output branches (see Fig. \ref{description}). Both the central leading pulse and the spatial repartition of the transmitted waves are shown to decay exponentially with the propagation distance and the distance to the central axis, respectively. This behavior is then studied on experimental realizations of this system and compared with theoretical predictions. The fast decay of the acoustic energy observed in these types of media suggests that ordered granular networks might be extremely relevant as impact mitigating materials.

Beyond the applications of these microstructured materials as acoustic shields, ordered networks of granular chains might serve as a model system to investigate the acoustic properties of natural granular piles, even though these are intrinsically disordered. Continuum mechanics approaches clearly fail to provide a reliable model of wave propagation in granular matters (see for example \cite{Makse}), due to the large force inhomogeneities present in stationary bead packs. Indeed, in disordered granular assemblies, acoustic waves are transmitted along force chains with preferential directions based on the inter-particle contact network resulting from the static force transmission \cite{Liu:1995}. A variety of studies have revealed this unique wave transmission pathway \cite{Bardenhagen:1998,Clark:2012,Roessig:2002,Owens:2011}. As a result, understanding wave propagation through networks of granular chains is the key to capture the acoustic behavior of granular matter, relevant in various applications of geophysics, soil mechanics, materials science, etc... Studies of wave propagation through y-branched granular chains \cite{Shukla,Daraio:2010,Ngo:2012,Qiong:2013} have provided a quantitative description of the elementary mechanism of wave splitting in granular packings. However, an understanding of the effective acoustic response of granular chain networks that involve multiple wave splittings, recombinations and reflections is still missing. The study of ordered granular arrays that allow for a rigorous treatment of the wave evolution aims at addressing such questions. The observations in uncompressed disordered packings of (i) an exponential decay of the wave amplitude with the distance of propagation \cite{Clark:2012,Owens:2011} (ii) and the spatio-temporal structure of the transmitted wave packet consisting of a leading pulse followed by a train of slower pulses \cite{Jia2}, similar to our observations on ordered arrangement of grains as reported in this study, suggest that both systems share strong similarities and support the relevance of our approach.

In the following, we present the experimental, numerical and theoretical tools used to investigate wave propagation in ordered networks of granular chains with different branching levels $N$ (Sections \ref{Experimental Setup}, \ref{Numerical Simulations} and \ref{Quasi-Particle Theory}, respectively). The next section presents the main features of wave propagation through ordered granular chain networks as observed in experiments and simulations, and compares these findings with theoretical predictions based on the description of the multiple wave splittings and recombinations in the network. Finally, the last section discusses the relevance of our findings for the design of efficient shielding metamaterials and for the understanding of acoustic properties of granular media. The approaches used in this paper are general, and could be used to design different stress wave-guiding materials, for example, to focus impact energy instead of spreading it. The good agreement between experiments and both numerical simulations and theoretical predictions of the nonlinear wave propagation validates the proposed approaches, and suggests that more complex material designs with intricate predetermined wave propagation pathways can be experimentally feasible.

\section{Experimental setup: assembled granular networks of various levels}

\label{Experimental Setup}
\begin{figure}[h]
\includegraphics[scale=1]{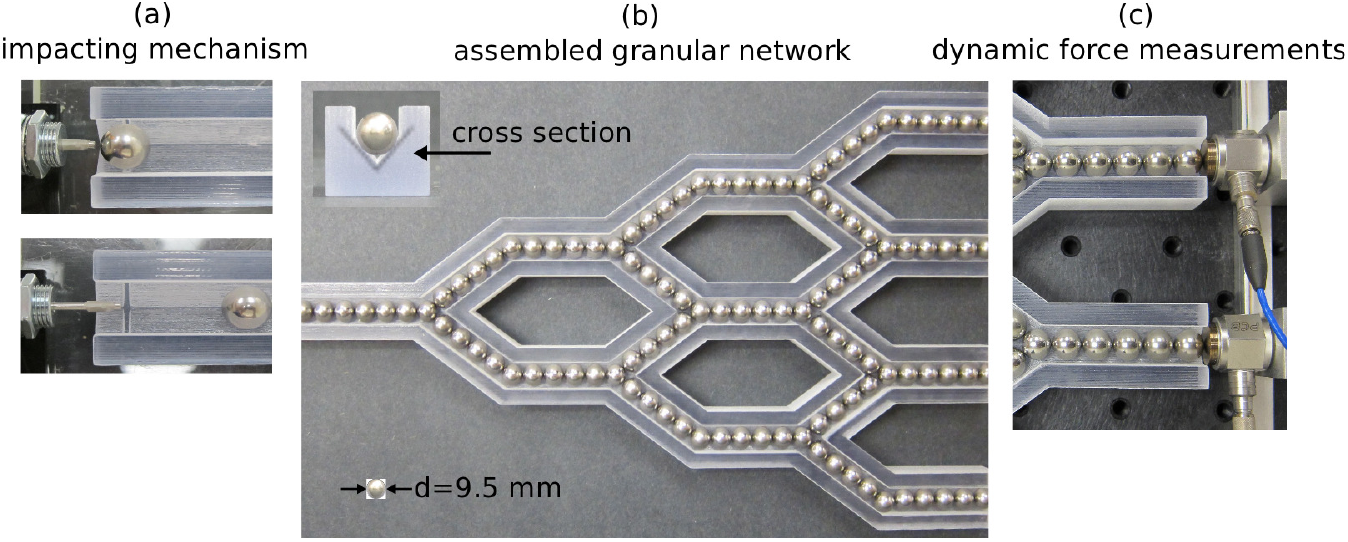}
\centering
\caption{Pictures of experimental setup. (a) Solenoid impacting mechanism. (b) Assembled granular network with inset showing the v-shaped cross section of the supporting channel structure. (c) Dynamic force measurements taken at each branch end.}\label{exp_setup}
\end{figure}

In order to study the effect of multiple wave splittings on the overall acoustic response of a granular medium, the granular network tested in our experiments is comprised of an initial segment that divides into two symmetric branches. Each new branch is then split into two symmetric branches, that can merge with other branches of the network (see Fig. \ref{description}). This process is repeated $N$ times, in order to produce a network of degree $N$ that consists of $(N+1)$ exit chains where the transmitted wave can be measured. Contrary to "classical" disordered granular media, the deterministic arrangement of granular chains allows for a rigorous description of the wave propagation and scattering within the network from which the effective acoustic response will be predicted (see Section \ref{Quasi-Particle Theory}). The initial segment consists of 10 spheres to allow an incident solitary wave to develop. The branching angle was chosen to minimize losses of the incident wave due to wave reflection, so that wave splitting remains the main mitigation mechanism. As the branch angle $\alpha$ increases, the loss around the corners continues to increase until no force is transmitted for $\alpha=90^{o}$ (see Section \ref{Quasi-Particle Theory}). However, a lower limit $\alpha=30^o$ exists, at which point the spheres on either side of the branching junctions come into contact. To guarantee clearance between neighboring particles at branch junctions within the assembly, $\alpha=35^o$ was chosen for all experimentally tested networks. In order to minimize dissipative losses along the length of the chains, each branching segment consisted of 6 spheres, which is sufficient to support the solitary wave length of approximately 5 particle diameters \cite{FNesterenko:2001}. The tested granular networks consist of spheres (precision ball bearings from McMaster-Carr) assembled in the supporting channel structures. The spheres are stainless steel (type 420C) with diameter, $D=9.5$ mm. The particle size and material used for experiments was chosen to simplify the experimental setup. However, the analysis described in Sections \ref{Quasi-Particle Theory} and \ref{Results} is valid for a homogeneous network of any size or material particles. The supporting channel structures are fabricated using VeroClear material with the Connex 500 3D printing system.

In order to investigate the effect of multiple splittings through a granular network, individual channel structures were printed for networks of various degrees, from $N=1$ up to $N=4$. Let us note that the case $N=1$ corresponding to y-branched granular systems have been investigated in several works \cite{Shukla,Daraio:2010,Ngo:2012,Qiong:2013}, providing a detailed description of the wave splitting mechanism. Precise alignment of the spheres in each network is crucial to ensure simultaneous arrival of pulses at branch junctions and observe recombinations of solitary waves in experiments. The supporting channel structures were fabricated with a v-shaped cross section (see the inset in Fig. \ref{exp_setup}(b)) to support and align the spheres and to reduce the number of contact points between the spheres and their support. A v-shaped cross section is less reliant on the exact dimensions of the supporting channel, since the particles will always rest at the same height along the centerline of the v, assuming a constant v-profile through the length. A square channel or a tube, for example, could have also been used instead of the v-shaped channel, however the dimension of these cross-sections would need to be very accurate: if too small, the particle motion would be severely impeded, and if too wide, the alignment of the sphere centers would be skewed. A slight tilt (approximately $2^o$) was induced on the experimental assembly to promote particle contacts in the network. Additionally, asymmetries within the granular network could arise by including sensor particles (similar to \cite{Daraio:2005}) within the branches. Therefore, force measurements were taken at each of the branch ends with piezoelectric dynamic force sensors (PCB 208C01 and PCB 208C02, with sensitivity $11.2 \, \mathrm{mV/N}$ and $112.4 \, \mathrm{mV/N}$, and an upper frequency response of 36kHz). A conditioner (PCB 481A02) amplified signals when necessary and the data was collected through a data acquisition board (NI BNC-2110 and NI PCI-6123).~The force between the last particle and a flat sensor will be higher than contact forces between particles in the chain, since the sphere-sensor stiffness is greater than the sphere-sphere stiffness. A previous study by Job et al. provides a general expression for the relative solitary wave amplitude in a chain of beads with respect to the amplitude between the last bead in contact with a wall, in terms of the bead and wall materials. Using the theoretical prediction described by \cite{Job:2005}, the arriving solitary wave amplitude in the branches contacting the sensors can be calculated as $F_{\mathrm{chain}}=F_{\mathrm{sensor}}/1.7$, for the stainless steel spheres and stainless steel impact cap used in experiments.

The assembled systems were excited by a single striker sphere, identical to the spheres in the network. The striker sphere was given an initial velocity with a solenoid mechanism (see Fig. \ref{exp_setup}(a)). To capture the variability between experiments, each network with $N=1$ to $N=4$ was disassembled and reassembled five times, with five impacts on each initial assembly. The incident solitary wave amplitude was determined to be $48.7 \mathrm{N} \pm2.2 \mathrm{N}$, based on repeated impacts of an $N=0$ system consisting of 10 spheres. This average impacting force corresponds to a striker velocity of $v_{\mathrm{striker}}=0.44 \, \mathrm{m/s}$, based on numerical simulations. However, the analysis performed in Sections \ref{Quasi-Particle Theory} and \ref{Results} is relevant for all impact forces, which are within the linear elastic response of the material. In experiments, it was advantageous to use a smaller impact force, to remain both within the operating frequency range of the force sensors, and within the yield limit of the material.

\section{Numerical approach: granular systems as networks of non-linear springs}
\label{Numerical Simulations}
In order to help the comparison between experimental observations and theoretical predictions, we use a discrete numerical model of wave propagation in granular networks. Numerous studies, such as those of \cite{Cundall:1979} and \cite{Sen:2008}, validated the use of a discrete element model to simulate the dynamic behavior of granular systems. The model considers particles as point masses connected by nonlinear, Hertzian springs \cite{Johnson:1987}. The repulsive force $F_{ij}$ between neighboring spheres $i$ and $j$ evolves as a power law $\delta_{ij}^{3/2}$ of their penetration distance. We used a Runge Kutta scheme to integrate the following system of equations, consisting of:

\begin{equation}\label{EOM}
m_{i}\ddot{\mathbf{u}}_{i}=\sum_{j=1}^{\textnormal{P}} \mathbf{F}_{ij} =\sum_{j=1}^{\textnormal{P}} K_{ij} \left|\boldsymbol{\delta}_{ij} \right|^{\nicefrac[]{3}{2}} \hat{\boldsymbol{\delta}}_{ij} ,
\end{equation}
\begin{equation*}
\boldsymbol{\delta}_{ij}= \left(\left(R_{i}+R_{j} \right) - \left| \mathbf{r}_{ij} \right| \right]) \hat{\mathbf{r}}_{ij}~~\textnormal{where}~~\mathbf{r}_{ij}=\left[\left(x_{j}-x_{i} \right)~\left(y_{j}-y_{i} \right)~\left(z_{j}-z_{i} \right) \right]
\end{equation*}
for each sphere $i$. In these Equations, $m$ is the particles mass, $\mathbf{u}_{i}=\left[u_{ix}~u_{iy}~u_{iz} \right]$ represents the particles $x$, $y$, and $z$ displacement from equilibrium and $P$ is the number of neighboring particles ($P$ equals two or three spheres plus two wall particles). $K_{ij}=\frac{4}{3}\left(\frac{1-\nu_{i}^{2}}{E_{i}}+\frac{1-\nu_{j}^{2}}{E_{j}}\right)^{-1}\left(\frac{R_{i}R_{j}}{R_{i}+R_{j}}\right)^{\nicefrac{1}{2}}$ is the contact stiffness between two particles $i$ and $j$ and $\boldsymbol{\delta}_{ij}$ is the penetration distance between two particles. $\boldsymbol{\delta}_{ij}$ takes a zero magnitude when particles $i$ and $j$ are not in contact. $E$ is the Young's Modulus, $\nu$ is the Poisson's ratio, and $R$ is the radius of the beads.

The discrete element simulations were performed in three-dimensional space, in order to accurately model the walls oriented at $45^{o}$ with respect to the horizontal plane. The walls were modeled as immovable particles of infinite radius, with the specified material properties. The average force on particle $i$ was calculated as $F_{i}=\left( \left(\frac{ \sum_{j}^{} \left|\mathbf{F}_{ij} \cdot \hat{\mathbf{x}}\right|}{2} \right)^{2} +\left(\frac{ \sum_{j}^{} \left|\mathbf{F}_{ij} \cdot \hat{\mathbf{y}}\right|}{2} \right)^{2} + \left(\frac{ \sum_{j}^{} \left|\mathbf{F}_{ij} \cdot \hat{\mathbf{z}}\right|}{2} \right)^{2} \right)^{\nicefrac{1}{2}}$ in numerical simulations to compare with the $F_{chain}$ force (calculated from $F_{sensor}$) measured in experiments.

The effects of gravity and dissipative losses present in experiments were omitted in the numerical simulations. The $~2$ degree tilt imposed in experiments corresponds to a force less than 0.1N at the end branch of the largest, N=4, tested network, which is sufficiently small compared to the measured signals to neglect this contribution (see \cite{Sen:2008}). While there are a number of proposed models for the dissipation of acoustic energy observed experimentally in an isolated chain (see, e.g., \cite{Coste:1997,Rosas:2007,Sen:2008,CarreteroGonzalez:2009,Herbold:2010}), the topic is still being explored. On this basis, we chose to neglect dissipation in the numerical simulations. As a result, wave splitting remains the dominant scattering mechanism.

For the simulations, we take a density {\bf $\rho=7800~\mathrm{kg/m^{3}}$}, Young's Modulus $E=200~\mathrm{GPa}$, and Poisson's ratio $\nu=0.28$ for the spheres, corresponding to the properties of the stainless steel beads used in the experiments (www.efunda.com). The printed VeroClear material, used for making the supporting channels, has a manufacturer specified density {\bf $\rho=1045~\mathrm{kg/m^3}$} and Young's Modulus $E= 2-3~\mathrm{GPa}$ (http://objet.com). To model the supporting channel material in numerical simulations, $E$ is taken as $2.5~\mathrm{GPa}$, and $\nu=0.35$ is assumed for the VeroClear material, which is within the normal range for polymers. Let us note that the numerical simulations were rather insensitive to the wall material properties.

\section{Theoretical approach: the quasi-particle theory}
\label{Quasi-Particle Theory}
The combination of discreteness and nonlinear contacts between particles causes excitations along uncompressed uniform chains of spheres to travel as single solitary waves or trains of solitary waves \cite{Nesterenko:1983,FNesterenko:2001,Job:2007,Sen:2008}. Individual solitary waves have specific properties determined by the size and material of the underlying particles, and each pulse travels unchanged along the length of a chain. Due to their nature, each solitary wave traveling through the chains can be modeled as a single particle with an equivalent momentum and kinetic energy. The effective or quasi-particles have a mass {\bf $m_{\mathrm{eff}}$} and velocity {\bf $V_{\mathrm{eff}}$} which are related to the mass $m$ of the individual spheres and the velocity $V_{\mathrm{SW}}$ of the solitary wave \cite{Job:2007,Ngo:2012} by

\begin{equation}
{m_{\mathrm{eff}}=1.345 \, m} ,
\end{equation}
\begin{equation}
{V_{\mathrm{eff}}=1.385 \left( \frac{\sqrt{5}}{2} \right)^{4} \left(\frac{ \pi \rho \left(1-\nu^{2} \right)}{2E} \right)^{2} V_{\mathrm{sw}}^{5}} .
\end{equation}

Using the quasi-particle approach has practical advantages for the network of granular chains studied here: it greatly simplifies the wave transmission calculations through bends and junctions by modeling them as a series of hard sphere collisions between quasi-particles, where the incident quasi-particle mass and velocity are always known. The velocity of the impacted quasi-particle is then calculated via conservation of linear momentum and energy, similar to the procedure described in \cite{Ngo:2012}. All particle masses, and thus effective particle masses, are identical for the present study. Utilizing the relationship between the effective quasi-particle velocity, solitary wave speed, and solitary wave force amplitude (leading to $V_{\mathrm{eff}}~\alpha~F_{\mathrm{sw}}^{\nicefrac{5}{6}}$), the transmission coefficients through the granular chain network can be derived.

\begin{figure}[h]
\includegraphics[scale=0.5]{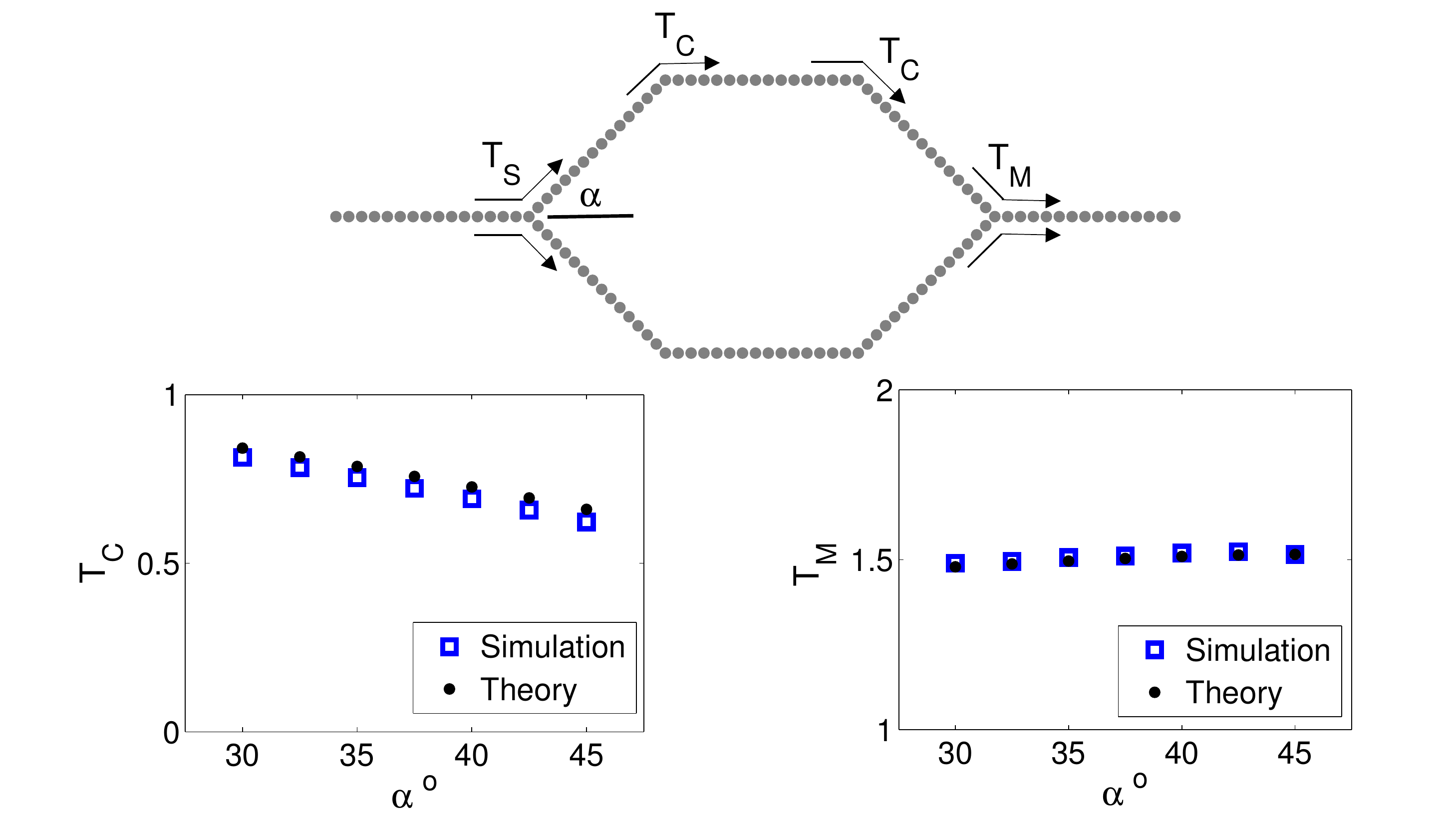}
\centering
\caption{ (Top) Schematic diagram of a unit cell configuration depicting the splitting transmission coefficient T\textsubscript{S}, the corner transmission coefficient T\textsubscript{C} and the recombination transmission coefficient T\textsubscript{M}. (Bottom) Comparison of numerical simulations and theoretical predictions for T\textsubscript{C} (Left) and T\textsubscript{M} (Right) for branch angles $\alpha$ between 30 and 45 degrees. In numerical simulations, the incident and transmitted amplitudes were calculated using the average maximum force from particles 7 through 11 along each segment.}\label{newcoeff}
\end{figure}

Wave propagation through the chain network involves three elementary mechanisms that are depicted in Fig. \ref{newcoeff} (Top): (1) pulse splitting (S) when a chain splits into two symmetric branches, (2) wave propagation through a chain with a corner (C), and (3) pulse recombination when two identical branches merge together (M). Their effect on the wave behavior can be fully described by introducing the transmission coefficients T\textsubscript{S}, T\textsubscript{C} and T\textsubscript{M}, respectively, that provide the ratio of the amplitude of the transmitted over the incident leading pulse through these junctions and corners.

\paragraph{Wave splitting} In a previous study, \cite{Ngo:2012} studied experimentally and numerically, the wave transmission along a chain that branches into two new chains. They confirmed the relevance of the quasi-particle approach to characterize wave splitting and showed that for a symmetric branching, as the one involved in our chain networks, the wave amplitude $F_{\mathrm{sw}}^{\mathrm{t}}$ in the two new branches is proportional to the incident wave amplitude ($F_{\mathrm{sw}}^{\mathrm{i}}$) and follows
\begin{equation}
\textnormal{T}_{\mathrm{S}}=\frac{F_{\mathrm{sw}}^{\mathrm{t}}}{F_{\mathrm{sw}}^{\mathrm{i}}}=\left(\frac{2 \cos \alpha}{2 \cos^{2} \alpha+1} \right)^{\nicefrac[]{6}{5}}. 
\label{Eq_Ts}
\end{equation}

\paragraph{Wave propagation through bend chains}
The transmitted amplitude through a corner with bend angle $\alpha$ can be described by considering a binary collision between the incident quasi-particle (initial velocity $V_{\mathrm{eff}}^{\mathrm{i}}$, and "post-collision" velocity $V_{\mathrm{eff}}^{\mathrm{r}}$) and the impacted quasi-particle ("post-collision" velocity $V_{\mathrm{eff}}^{\mathrm{t}}$). Assuming complete transfer of linear momentum along an axis parallel to the transmission branch through the bend, we obtain the following expression: $m_{\mathrm{eff}}V_{\mathrm{eff}}^{\mathrm{i}} \cos{\alpha}=m_{\mathrm{eff}}V_{\mathrm{eff}}^{\mathrm{t}}$. Solving for $\displaystyle \frac{V_{\mathrm{eff}}^{\mathrm{t}}}{V_{\mathrm{\mathrm{eff}}}^{i}}$ and relating the effective quasi-particle velocity to the solitary wave force amplitude using $\displaystyle \frac{V_{\mathrm{eff}}^{\mathrm{t}}}{V_{\mathrm{eff}}^{\mathrm{i}}}=\left(\frac{F_{\mathrm{sw}}^{\mathrm{t}}}{F_{\mathrm{sw}}^{\mathrm{i}}} \right)^{\nicefrac[]{6}{5}}$, the resulting transmission coefficient through a corner is:
\begin{equation}
\textnormal{T}_{\mathrm{C}}=\frac{F_{\mathrm{sw}}^{\mathrm{t}}}{F_{\mathrm{sw}}^{\mathrm{i}}}=\left(\cos \alpha \right)^{\nicefrac[]{6}{5}} .
\label{Eq_Tc}
\end{equation}
During wave propagation through bends in the physical system, the corner particle deforms into the wall and subsequently rebounds, sending a reflected pulse back through the incident branch. The numerical simulations take into account the wall material properties, and were performed for a range of wall stiffness, or Young's moduli. We observed that changing the wall stiffness primarily effects how quickly the reflected pulse is sent back through the incident chain, and has a negligible effect on the transmission coefficient $\mathrm{T}_{\mathrm{C}}$. Under this assumption, the energy lost by the leading pulse through a corner, i.e. reflected, is related to the component of the incident particle velocity perpendicular to the transmission branch, before and after the binary quasi-particle collision, given by $V_{\mathrm{eff}}^{i} \sin \alpha$. As a result, this energy lost through branch bends is $\frac{1}{2}m_{\mathrm{eff}} \left(V_{\mathrm{eff}}^{\mathrm{r}} \right)^{2}$, leading to a ratio of energy lost by the leading pulse over the incident energy $\displaystyle \frac{E_{\mathrm{d}}}{E^{\mathrm{i}}} = \sin^2\alpha$.

\begin{figure}[h]
\includegraphics[scale=0.5]{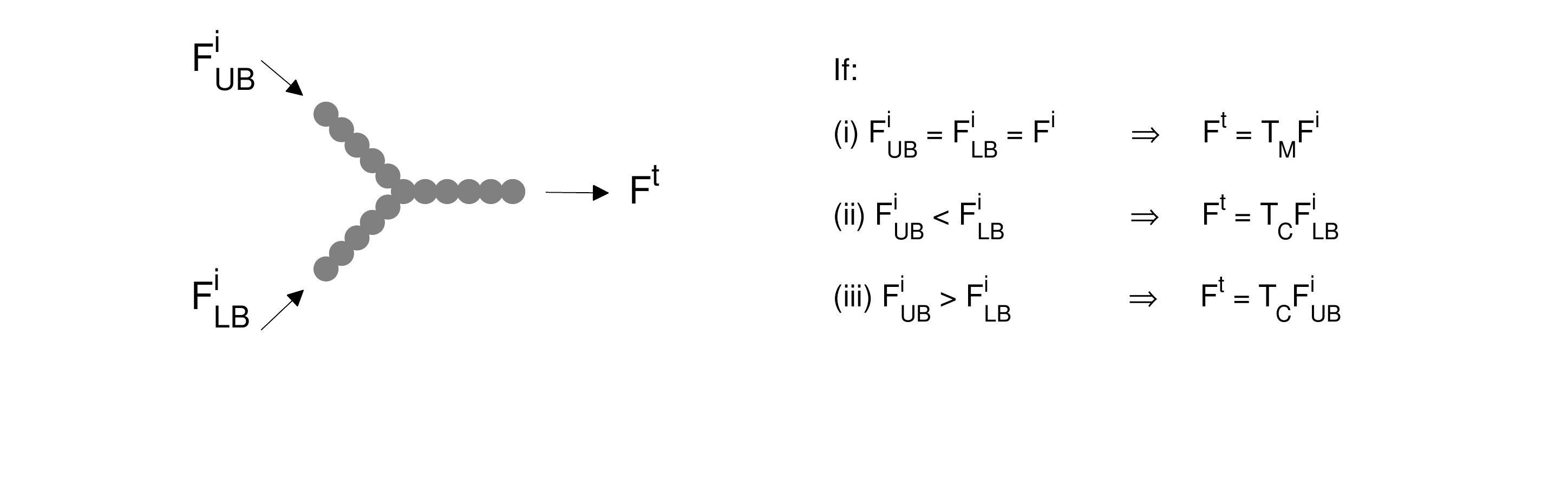}
\centering
\caption{ Schematic diagram describing the choice of transmission coefficient used to model the transmitted leading pulse amplitude $\mathrm{F}^{\mathrm{t}}$ at junctions where two granular chains merge together. $F_{\mathrm{LB}}^{\mathrm{i}}$ and $F_{\mathrm{UB}}^{\mathrm{i}}$ denote the incident force amplitudes in the lower and upper branch, respectively.}\label{MvsC}
\end{figure}

\paragraph{Wave recombination}
The recombination junctions are modeled as two incident quasi-particles symmetrically impacting a single transmission quasi-particle. It is assumed that the incident, or impacting spheres, travel after the collision with a reflected velocity coincident with the impact velocity, a physical constraint imposed by the supporting channel system. Momentum along y is conserved by symmetry, and conservation of momentum along x gives the following relation between the effective velocity $V_{\mathrm{eff}}^{\mathrm{i}}$ of the incident quasi-particles, the reflected velocity $V_{\mathrm{eff}}^{\mathrm{r}}$ of the incident quasi-particles, and the transmitted velocity $V_{\mathrm{eff}}^{\mathrm{t}}$ of the impacted quasi-particle, leading to $2m_{\mathrm{eff}}V_{\mathrm{eff}}^{\mathrm{i}} \cos{\alpha}=2m_{\mathrm{eff}}V_{\mathrm{eff}}^{\mathrm{r}} \cos{\alpha}+m_{\mathrm{eff}}V_{\mathrm{eff}}^{\mathrm{t}}$. Conservation of energy during the collision between the three particles is expressed as $2\frac{1}{2}m_{\mathrm{eff}} \left(V_{\mathrm{eff}}^{\mathrm{i}} \right)^{2}=2\frac{1}{2}m_{\mathrm{eff}} \left(V_{\mathrm{eff}}^{\mathrm{r}} \right)^{2}+\frac{1}{2}m_{\mathrm{eff}} \left(V_{\mathrm{eff}}^{\mathrm{t}} \right)^{2}$. Using $\displaystyle \frac{V_{\mathrm{eff}}^{\mathrm{t}}}{V_{\mathrm{eff}}^{\mathrm{i}}}=\left(\frac{F^{\mathrm{t}}_{\mathrm{sw}}}{F^{\mathrm{i}}_{\mathrm{sw}}} \right)^{\nicefrac[]{6}{5}}$, the transmission coefficient when pulse recombination is involved is
\begin{equation}
\textnormal{T}_{\mathrm{M}}=\frac{F^{\mathrm{t}}_{\mathrm{sw}}}{F^{\mathrm{i}}_{\mathrm{sw}}}=\left(\frac{4 \cos \alpha}{2 \cos^{2} \alpha+1} \right)^{\nicefrac[]{6}{5}}.
\label{Eq_Tr}
\end{equation}
In order for pulses to recombine, the incident pulses in the upper and lower branches must have identical wave structures and arrive at a chains junction simultaneously. At some chain merging locations, the leading pulses traveling in the upper and lower incident branches will arrive with unequal amplitudes (and at separate times) as a result of differing propagation path histories within the granular chain network. In this case there will be multiple transmitted pulses, a larger primary pulse and a smaller amplitude trailing pulse. Therefore, not all chain merging locations result in pulse recombination. To describe the leading pulse amplitude transmission, chain merging junctions are modeled either as recombination junctions or as corners, depending on the incident upper and lower branch wave structure, as detailed on Fig. \ref{MvsC}.

For $\alpha<45^{o}$, we observed a reflected wave at the splitting junctions and corners, but not at the recombination junctions. For both the theoretical quasi-particle predictions and numerical simulations, the incident hard spheres at recombination junctions were observed to possess a forward velocity after the collision, relative to the direction of their initial velocity. The continued forward momentum of the impacting effective particles produced a secondary solitary wave train \cite{Job:2007} in the recombination transmission branch. The amplitude of the secondary transmitted solitary wave decreases with increasing $\alpha$. The amplitude of the secondary wave train is small with respect to the leading pulse, and we will focus on the latter in the remainder of this study. 

The theoretical predictions for T\textsubscript{C} and T\textsubscript{M} are in excellent agreement with those obtained from numerical simulations for branch angles between $30^{o}$ and $45^{o}$, as shown in Fig. \ref{newcoeff}. It should be noted that the quasi-particle coefficients could also be used to track the amplitude of the trailing or secondary pulses traveling through the network. However, for the purpose of this study we focus on the distribution of the leading, i.e. largest, amplitude pulses. The expressions for the transmission coefficients given by Eqs. (\ref{Eq_Ts}), (\ref{Eq_Tc}) and (\ref{Eq_Tr}) serve now as elementary bricks to predict the overall acoustic behavior of the networks of granular chains.

\section{Results}
\label{Results}

\begin{figure}[!ht]
\includegraphics[scale=0.60]{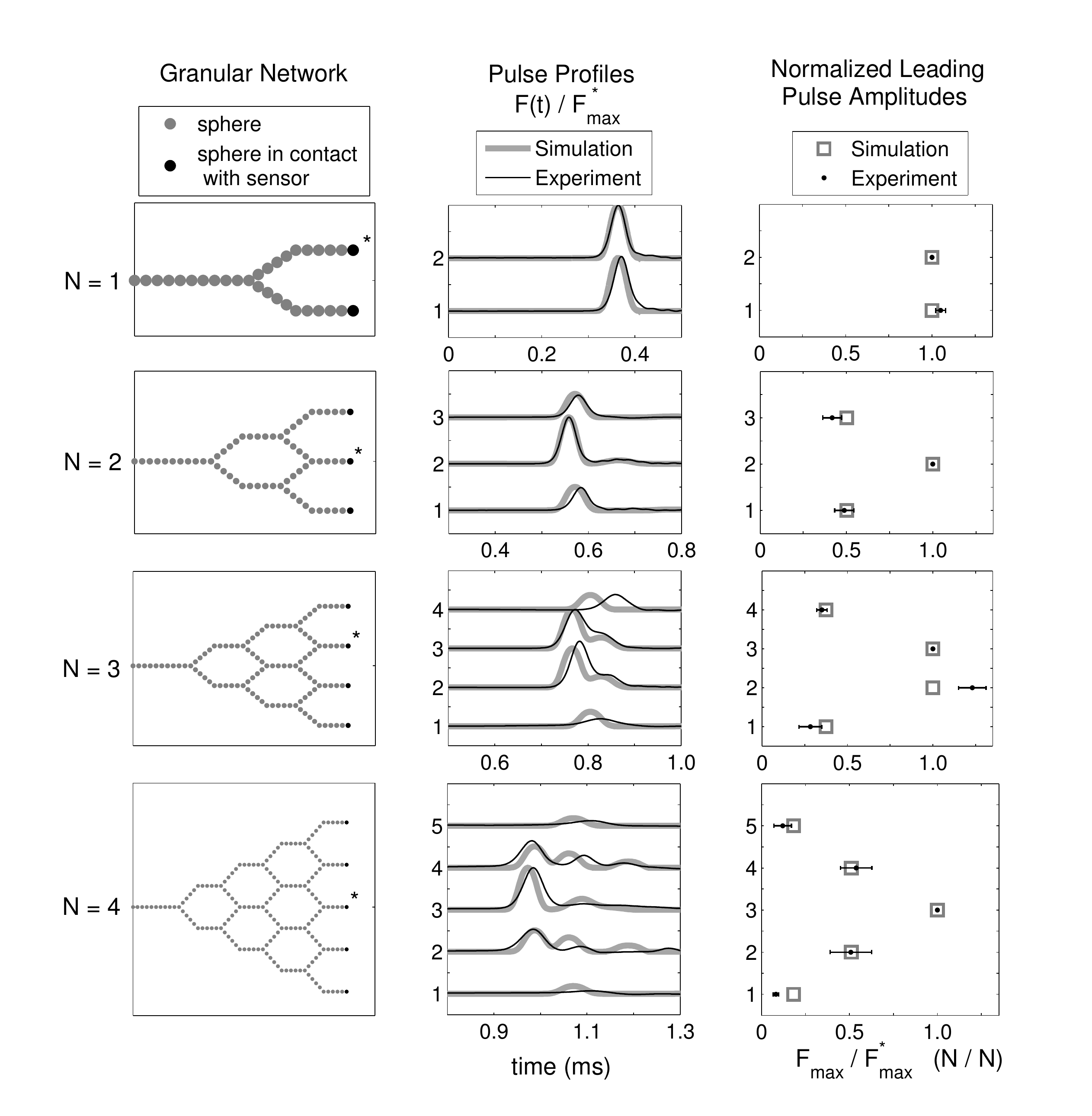}
\centering
\caption{Numerical simulations and experimental results of the forces at each branch end for the tested N=1 through N=4 networks. (Left) Schematic diagram of the branch network for each level. The force profiles correspond to particles in contact with the sensors in experiments, denoted in black. For each level, the amplitude was normalized by the amplitude of the particle indicated with a $^{*}$, referred to as $\mathrm{F}_{\mathrm{max}}^{*}$. (Middle) Numerical simulations, solid grey line, and a single experimental test, representative of the mean value shown in the right panel, solid black line, for each of the tested networks, N=1-4. The force profiles of each branch end are offset by $1~\nicefrac[]{N}{N}$ for visual clarity. The zero time indicates the time of impact in numerical simulations. For comparison, the experimental zero time was shifted to coincide with the arrival time of the force at the particle indicated with a $^*$. (Right) Distribution of the leading pulse amplitudes at the branch ends, normalized by $\mathrm{F}_{\mathrm{max}}^{*}$, for each branch level N. The grey boxes indicate results from numerical simulations, and the black dots and error bars represent the mean and standard deviation from experiments.
}\label{numexpprofile}
\end{figure}

Experiments and numerical simulations are performed for each of the $N=1$ up to $N=4$ granular networks described in Section \ref{Experimental Setup} (also pictured in Fig. \ref{numexpprofile} (Left) ). In numerical simulations, the last branch segments were extended by five particles to obtain an average transmitted amplitude through the exit channel corresponding to the end particle in experiments.  Figure \ref{numexpprofile} depicts the normalized force profiles at each of the $N+1$ exit branches in both numerical simulations and a single experiment (middle panel) from each of the $N=1-4$ networks. While exact force profiles varied slightly between the five experiments on each $N$ network due to slightly different arrangements of particles within the channels, the general features including the number of pulses arriving at the branch ends and the relative amplitudes of each pulse were consistent over all experiments. The left panel of Fig. \ref{numexpprofile} indicates the variability of the relative amplitudes observed in experiments between the five assemblies and five impacts on each network. The small trailing oscillations observable in the $N=1$ and $2$ experiments are the result of the last sphere-sensor interaction, and are not an inherent feature of the network structure.\footnote{The larger amplitude pulses that cross the $N=1$ and $2$ networks correspond to shorter temporal wavelengths, resulting in higher frequency waves which slightly resonate with the sensors.} Overall, experiments and simulations are in good qualitative agreement for all of the experimentally tested networks.

\subsection{Description of the central largest leading pulse}
\begin{figure}[!ht]
\includegraphics[scale=0.33]{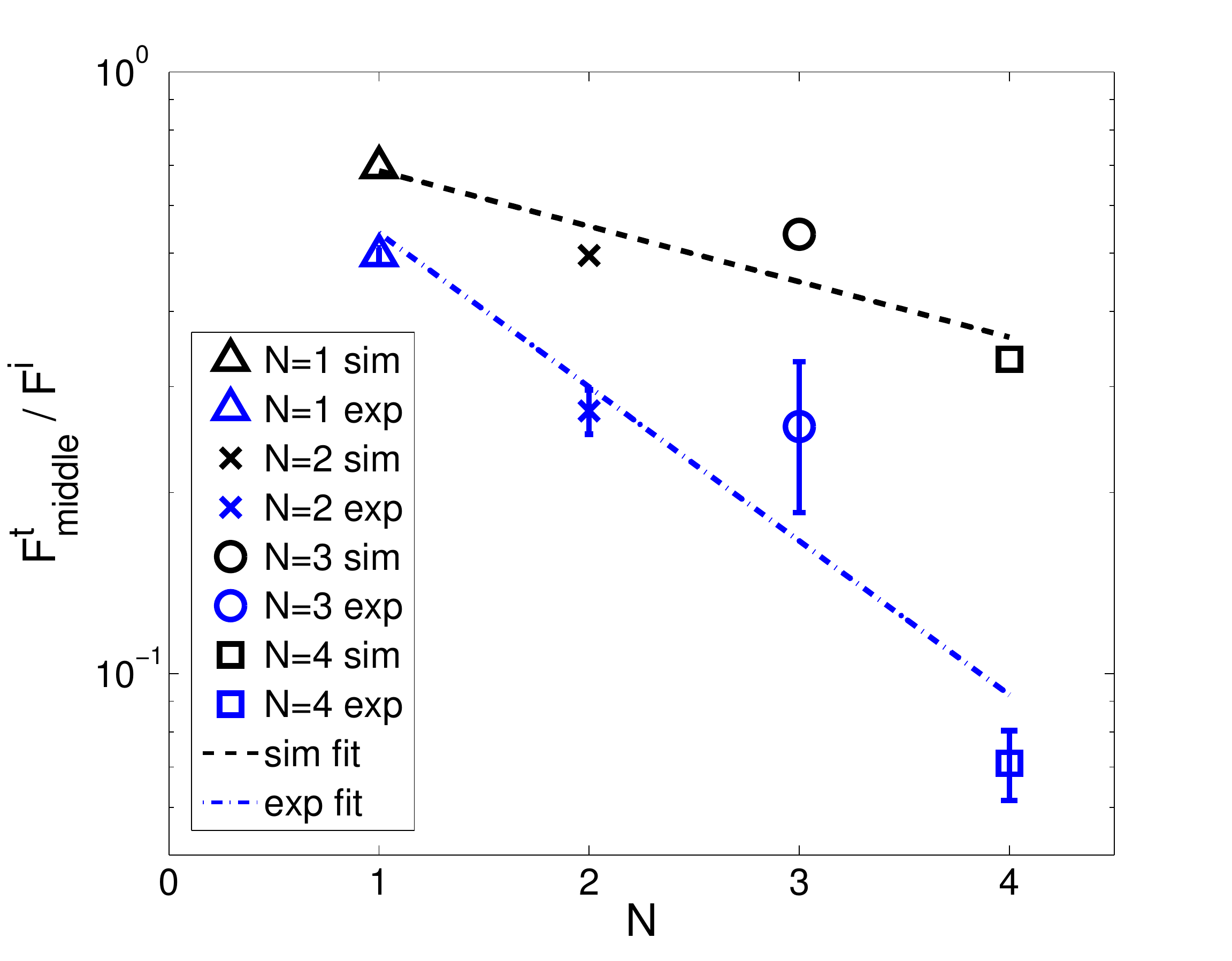}
\includegraphics[scale=0.33]{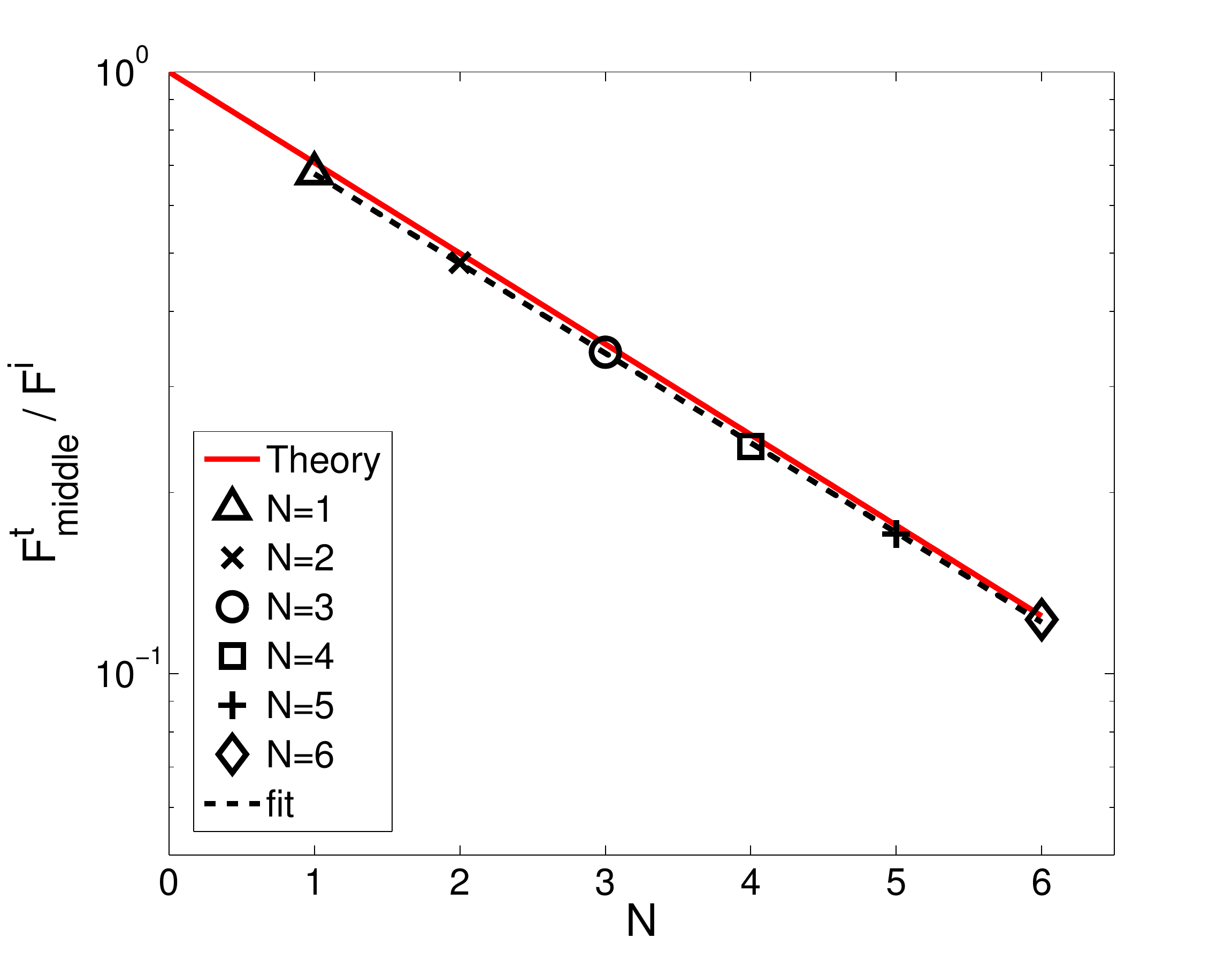}
\centering
\caption{ $F_{\mathrm{middle}}^{\mathrm{t}}$(normalized by the input force, $\mathrm{F}^{\mathrm{i}}$) as a function of branch levels, $N$. (Left) Numerical simulations are represented in black markers and experimental results in blue markers for $N=1-4$ for the tested network with six particles per chain. The linear fit through the simulation data is represented by a dotted black line and the linear fit through the experimental data by a dot-dashed blue line. (Right) Numerical simulations are represented in black markers for $N=1-6$ for the network with 16 particles per chain. The linear fit through the simulation data is represented by a dotted black line and the theoretical predictions by a solid red line.}\label{Fmiddle_vs_N}
\end{figure}

Experiments and numerical simulations are compared more quantitatively on the left panel of Fig. \ref{Fmiddle_vs_N} where the transmitted wave amplitude $F_{\mathrm{middle}}^{\mathrm{t}}$, defined as $F^{t}(N,r=0)$, in the middle branch is represented as a function of the network level $N$. For $N$ even networks, the force is directly measured on the central branch (marked by a star sign on the left panel of Fig. \ref{numexpprofile}), while for $N$ odd, this value is extrapolated from the transmitted amplitude measured in the other exit chains according to a procedure described below. For experiments and numerical simulations, an exponential law can be used to describe the decay of the transmitted wave amplitude in the central branch as a function of the level $N$ of the network:
\begin{equation}
\displaystyle F_{\mathrm{middle}}^{\mathrm{t}} = F^{\mathrm{i}} e^{-\frac{N}{N_0}} 
\end{equation}
where the actual value of $N_0$ is obtained from the exponential fit of the data shown in Fig. \ref{Fmiddle_vs_N} and listed in Table \ref{table:fittingvalues}. Here, $F^{\mathrm{i}}$ is the force amplitude of the incident wave generated by the impact of the striker. This behavior means that granular chain networks are very efficient for mitigating waves, and systems of size $N \gg N_0$ will be preferentially chosen for designing strongly mitigating metamaterials. In experiments, the effects of dissipation in isolated chains is evident and results in wave amplitude decay through each chain \cite{Rosas:2007,Sen:2008,CarreteroGonzalez:2009,Herbold:2010}, and consequently a lower value of $N_{0}$ compared to the conservative simulations. However, the description of $F_{\mathrm{middle}}^{\mathrm{t}}$ by an exponential decay remains fairly good.

To understand such a property of the chain network, we use the theoretical description proposed in Section \ref{Quasi-Particle Theory} to predict the effective acoustic behavior of the network. 
As illustrated in Fig. \ref{description}, one can predict the property of the transmitted wave by following its path though the network from the initial chain until the central exit chain. Since the wave velocity decays with its amplitude, only some of the various paths linking the entry to the exit of the network will contribute to the central leading pulse. Here, the largest pulses will result from the propagation through the central chains of the network. For example, for $N = 2$, the central transmitted pulse is the result of two wave splitting processes, the propagation of the wave through one corner, and one wave recombination, leading to $\displaystyle F_{\mathrm{middle}}^{\mathrm{t-theory}}(N = 2) = F^{\mathrm{i}} T_{\mathrm{S}}^2 T_{\mathrm{C}} T_{\mathrm{M}}$. To move from $N = 2$ to $N = 4$ \--- as from any network of level $N$ to another one of level $N+2$ \--- the wave will go through the same four processes again, leading to the predictions
\begin{equation}
\left\{
\begin{array}{l c c}
\displaystyle F_{\mathrm{max}}^{\mathrm{t-theory}}(N) & = & F^{\mathrm{i}} (T_{\mathrm{S}})^N (T_{\mathrm{C}})^{\nicefrac{N}{2}} \left ( T_{\mathrm{M}} \right )^{\nicefrac{N}{2}} \quad \mathrm{for \, } N \mathrm{\, even \, networks}
\vspace{10pt}
\\
\displaystyle F_{\mathrm{max}}^{\mathrm{t-theory}}(N) & = & F^{\mathrm{i}} (T_{\mathrm{S}})^N (T_{\mathrm{C}})^{\frac{N+1}{2}} (T_{\mathrm{M}})^{\frac{N-1}{2}} \quad \mathrm{for \, } N \mathrm{\, odd \, networks}
\label{Eq_Fmiddle}
\end{array}
\right .
\end{equation}
For $N$ even, $F_{\mathrm{max}}^{\mathrm{t}}=F_{\mathrm{middle}}^{\mathrm{t}}$. Rearranging Eq. (\ref{Eq_Fmiddle}) and using the expressions of the transmission coefficients $T_{\mathrm{S}}$, $T_{\mathrm{C}}$ and $T_{\mathrm{M}}$ given in Section \ref{Quasi-Particle Theory}, one can predict the amplitude of the largest transmitted pulse for a chain network with even $N$
\begin{equation}
\displaystyle F_{\mathrm{middle}}^{\mathrm{t-theory}}(N) = F^{\mathrm{i}} e^{-\frac{N}{N_0^{\mathrm{th}}}} \quad \mathrm{with} \quad N_0^{\mathrm{th}} = \frac{5}{12 \log \left(\frac{\left ( 1 + 2 \cos^2\alpha \right )^{\nicefrac{3}{4}}} {2 \cos \alpha} \right )}
\label{Eq_N0}
\end{equation}
This calculation for even values of network degree $N$ results in the same exponential decay of the central pulse amplitude for $N$ odd, after extrapolating $F_{\mathrm{middle}}^{\mathrm{t}}$from the distribution of leading pulse amplitudes along $r$.

Our theoretical approach captures the exponential behavior of the central leading pulse observed in the granular chain network, suggesting that wave splitting in the network is here the dominant mechanism. However, the theoretical value of $N_0^{\mathrm{th}} = 2.88$ for the branching angle $\alpha = 35^o$ used in the experimental and numerical networks, with 6 particle chain lengths, compares rather poorly with the measured decay rate $N_0$ given in Table \ref{table:fittingvalues}. This discrepancy comes from the limited number of particles in each chain, while our theory assumes long chains. For long chains, pulses of different amplitudes propagating in the granular network will have sufficient propagation distance to separate and arrive at different times at the next junction since the the solitary wave velocity is a slowly increasing function of its amplitude, with $V_{SW}~\alpha~F_{SW}^{\nicefrac{1}{6}}$. For the six particle chains used in both the experiments and the numerical simulations, the arrival time of waves at chain merging junctions is not sufficiently different to avoid some overlap between the primary pulse and trailing pulses. The pulse overlapping is most clearly seen in the pulse structure reaching the central exit branches in the $N=3$ network shown in Fig. \ref{numexpprofile}. As a result, the transmitted amplitude is slightly underestimated in our model, and the measured value for $N_0$ is larger than the predicted one.

To test this idea, we simulated wave propagation in a network with a larger number of particles in each chain. The amplitude of the central transmitted pulse is shown in the right panel of Fig. \ref{Fmiddle_vs_N} as a function of the system size, for network degrees up to $N=6$ and individual chains consisting of 16 particles. Our observations for these numerical simulations of long branch network ($N_0 = 2.9$) are in excellent agreement with the theoretical exponential decay ($N_0^{\mathrm{th}} = 2.88$).

\subsection{Description of the spatial repartition of transmitted leading pulses}
\begin{figure}[!ht]
\includegraphics[scale=0.33]{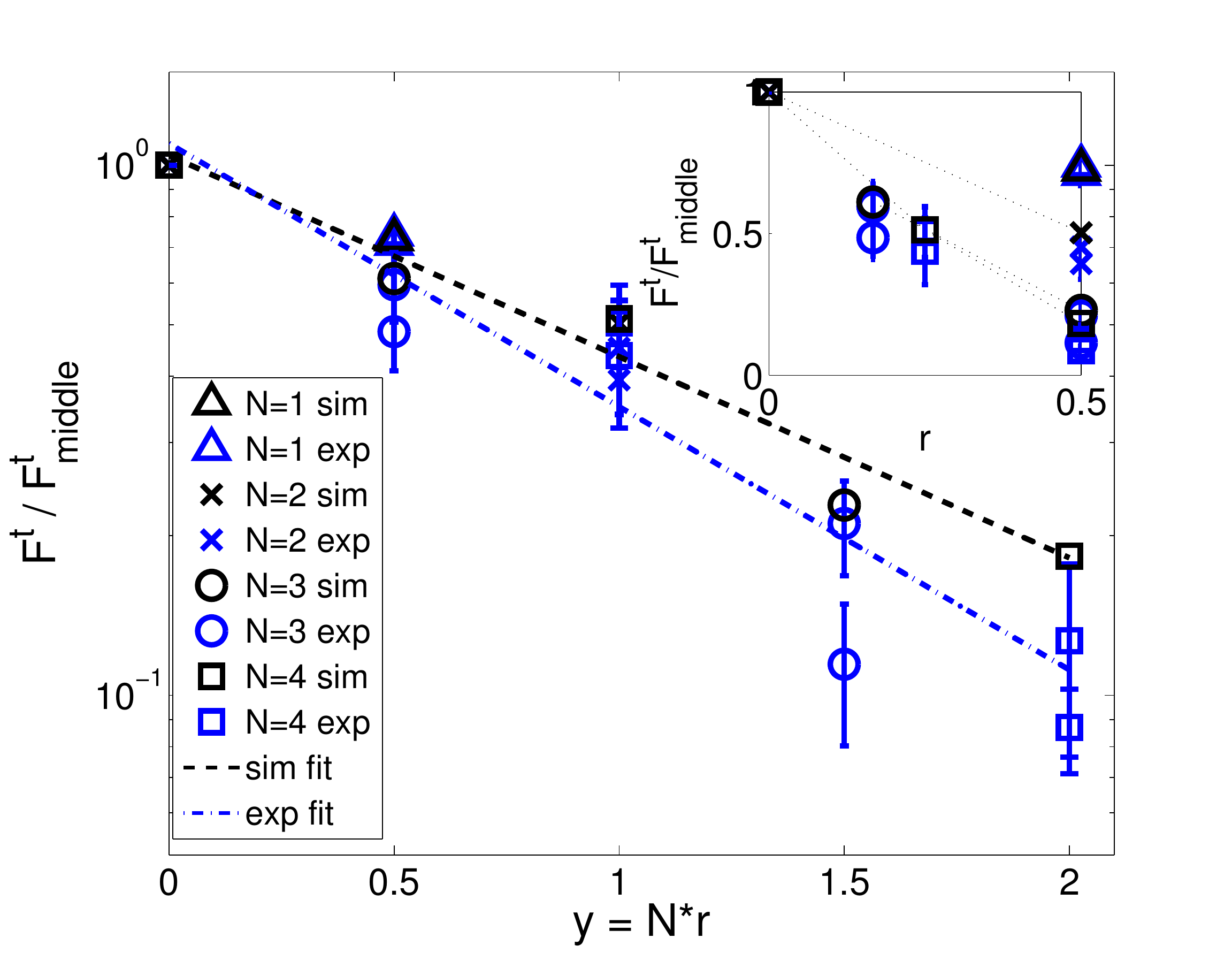}
\includegraphics[scale=0.33]{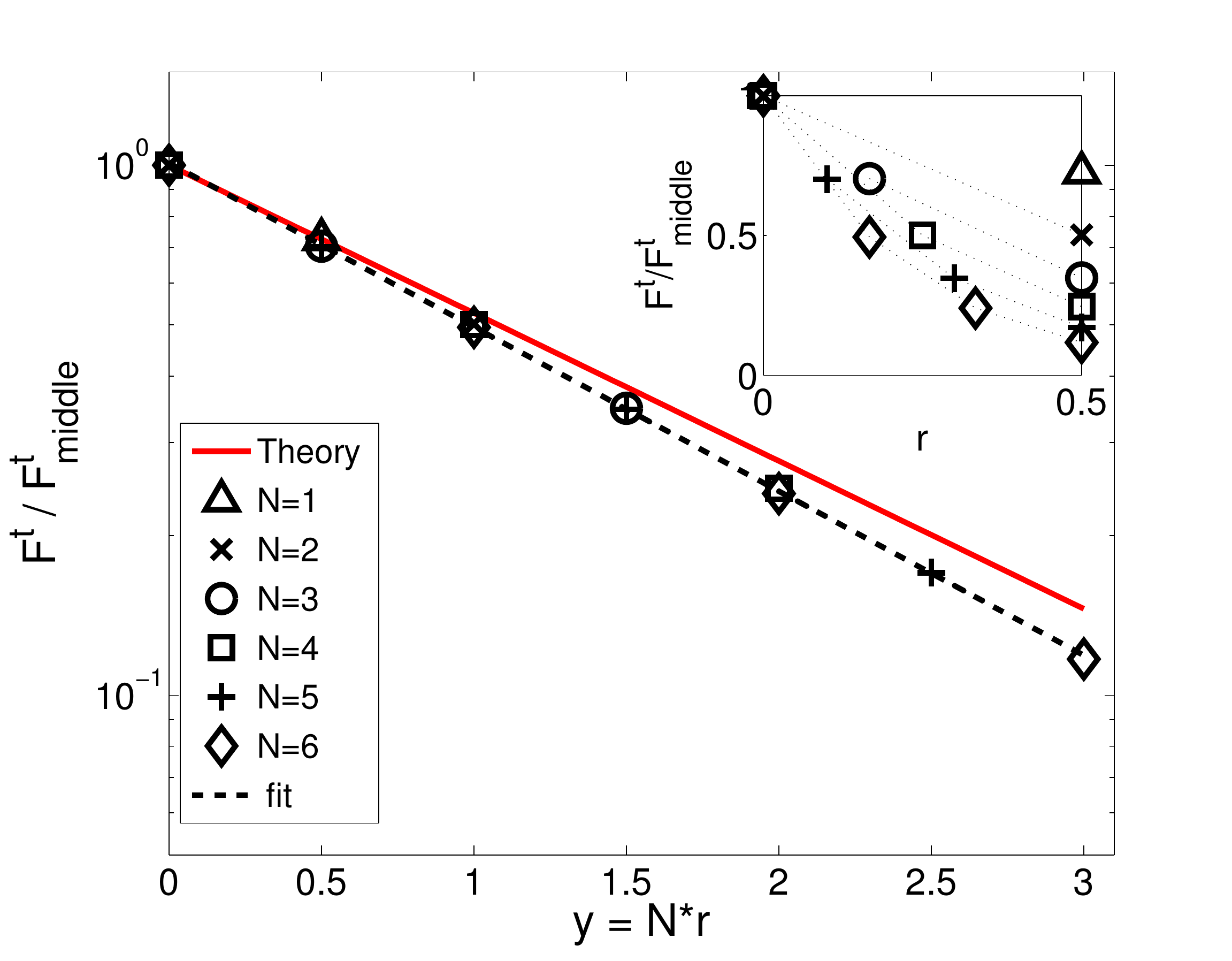}
\centering
\caption{Spatial repartition of the normalized transmitted force $F^{\mathrm{t}}/F_{\mathrm{middle}}^{\mathrm{t}}$ as a function of the distance $y = N \, r$ to the middle axis of the network. The insets show the force spatial distribution $F^{\mathrm{t}}/F_{\mathrm{middle}}^{\mathrm{t}}$ as a function of the normalized distance $r$. (Left) Numerical simulations are represented in black markers and experimental results in blue markers for $N=1-4$ for networks with six particles per chain. The linear fit through the simulation data is represented by a dotted black line and the linear fit through the experimental data by a dot-dashed blue line. (Right) Numerical simulations are represented in black markers for $N=1-6$ for networks with 16 particles per chain. The linear fit through the simulation data is represented by a dotted black line and the theoretical predictions by a solid red line.}\label{F_vs_y}
\end{figure}

We now go beyond the wave transmission in the middle exit chain, and study the full distribution of transmitted wave amplitudes as a function of the normalized distance $r$ to the middle axis of the network (see Fig. \ref{description}). Here, $r=0$ corresponds to the middle exit chain while $r = 0.5$ and $r = -0.5$ correspond to the side exit chains. For symmetry reasons, we will focus only on the values of the transmitted amplitude $F^{\mathrm{t}}(r)$ for positive values of $r$. The spatial repartition of the leading pulses are shown in the inset of Fig. \ref{F_vs_y}(left) for various network sizes $N$ for both simulations and experiments. After normalizing the ordinate by the central (largest) leading pulse amplitude $F^{\mathrm{t}}_{\mathrm{middle}}$ and multiplying the abscissa by the network degree $N$, all data collapse onto two master curves: (1) experiments and (2) simulations, as shown on the main panel of Fig. \ref{F_vs_y}(left). The linear variation in this semi-logarithmic representation suggests
\begin{equation}
\displaystyle F^{\mathrm{t}}(r) = F^{\mathrm{t}}_{\mathrm{middle}} \, e^{-\frac{N r}{\xi_0}}
\end{equation}
The values of $\xi_0$ obtained from the fit of the data by an exponential spatial distribution of transmitted amplitudes are given in Table \ref{table:fittingvalues}. Our numerical model captures reasonably well the spatial distribution observed in experiments as well as the value of $\xi_0$, despite the important role played by dissipation, as illustrated by the mitigation of the largest pulse through the network shown in Fig. \ref{Fmiddle_vs_N}. The normalization by $F_{\mathrm{middle}}^{\mathrm{t}}$ significantly reduces the effect of dissipation present in the experiments.
The network structure not only leads to an efficient mitigation of the incident wave along $r=0$, but also to a rapid decrease of the wave amplitude along an axis perpendicular to the line of impact. Taking now $y = N \, r $ as the rescaled distance of the exit chain to the central one, the spatial distribution of the leading pulses $F^{\mathrm{t}}(y) = F_{\mathrm{middle}}^{\mathrm{t}} e^{-\frac{y}{\xi_0}}$ indicates that the acoustic energy decreases exponentially rapidly as one moves away from the center of the network.

To understand such a strong directionality, we count the number of wave splittings, recombinations and corners along the path followed by the leading pulse reaching the branch at distance $r$ from the middle chain (see Fig. \ref{description}), and describe the wave propagation using the three transmission coefficients introduced in Section \ref{Quasi-Particle Theory}. For chains with either even or odd levels $N$, one obtains
\begin{equation}
F^{\mathrm{t-theory}}(r) = F^{\mathrm{i}} \left [ T_{\mathrm{S}} \left (T_{\mathrm{C}} \right )^{\frac{1}{2} + r} \left (T_{\mathrm{M}} \right )^{\frac{1}{2} - r} \right ]^N
\end{equation}
Let us note that $r$ takes a finite number of values, with $\displaystyle r \in \{0, \, \frac{1}{N}, \, ... \, , \, \frac{1}{2} \}$ and $\displaystyle r \in \{\frac{1}{2N}, \, \frac{3}{2N} , \, ... \, , \, \frac{1}{2} \}$ for even and odd network degrees, respectively. Using the expressions of the different transmission coefficients as a function of the branch angle $\alpha$ given in Eqs. (\ref{Eq_Ts}), (\ref{Eq_Tc}) and (\ref{Eq_Tr}), the spatial distribution of the transmitted wave can be expressed as
\begin{equation}
F^{\mathrm{t-theory}}(r) = F^{\mathrm{t-theory}}_{\mathrm{middle}} \, e^{-\frac{y}{\xi_0^{\mathrm{th}}}} \quad \mathrm{with} \quad \xi_0^{\mathrm{th}} = \frac{5}{12 \log \left(\frac{2}{\sqrt{1+2 \cos^2 \alpha}} \right )}
\label{Eq_F(r)}
\end{equation}
Let us note that $ \frac{3}{2} $ provides a rough estimate of the decreasing length scale of the wave spatial structure since $1.2 < \xi_0^{\mathrm{th}} < 1.8 $ for the range of admissible angles $30^o \leq \alpha \leq 45^o$.

The predicted value $\xi_0^{\mathrm{th}} = 1.57 $ for the angle used in the simulations and experiments overestimates the observed values given in Table \ref{table:fittingvalues}. Here again, overlap of primary and trailing pulses occur at chain merging junctions with different amplitude waves in the upper and lower branches for the short (6 particles) branch networks. This overlap is not accounted for in our theoretical model which simplifies these unsymmetrical chains junctions as corners, resulting in an underestimation of the transmitted amplitude at several exit branches within the network. The neglected pulse overlapping is more prevalent for central exit branches, {\it i.e.} for small values of $r$. In order to obtain a better comparison with the theoretical predictions, we performed simulations of wave propagation for long branch networks, which allow sufficient time for waves taking different and non-equivalent paths to separate and avoid solitary wave overlapping. Spatial distributions of transmitted amplitudes for the 16 particle chain networks with $N=1$ up to $N=6$ are represented on the right panel of Fig. \ref{F_vs_y}. Once renormalized by $F^{\mathrm{t}}_{\mathrm{middle}}$, and plotted as a function of the distance $y = r \, N $, the numerical results compare well with the predicted spatial distribution given in (Eq. \ref{Eq_F(r)}).

\begin{table}[!ht]
\caption{Comparison of the decay length scales $N_{0}$ and $\xi_{0}$ for theoretical predictions, long branch network (16 particles per branch) numerical simulations and short branch network (6 particles per branch) experiments and numerical simulations.}
\centering
\begin{tabular}{c c c c c}
\hline\hline
~~ & Theory & Long Branch & Short Branch & Short Branch \\ [0.5ex] 
~~ & ~ & Simulation & Simulation & Experiment \\ [0.5ex]
\hline
$N_{0}$ & 2.88 & 2.9 & 4.7 & 1.7 \\
$\xi_{0}$ & 1.57 & 1.4 & 1.1 & 0.9 \\[1ex]
\hline
\end{tabular}
\label{table:fittingvalues}
\end{table}
Table \ref{table:fittingvalues} summarizes the values of $\xi_{0}$ and $N_{0}$ obtained from experiments and numerical simulations of the short branch tested networks, as well as the numerical simulations performed for the long branch networks. For $N>1$ odd, the $F^{\mathrm{t}}_{\mathrm{middle}}$ value was obtained through linear regression of the spatial distribution of transmitted amplitudes on the $\log{(F^{\mathrm{t}})}$ vs $r$ data. For experiments, the linear regression is performed for both sides ($\pm r$) of the network in order to achieve a better precision on $\xi_0$ and estimate the error bar. For $N=1$, we use the theoretical expression of $\xi_{0}^{\mathrm{th}}$ of Eq. (\ref{Eq_F(r)}) to get an extrapolated value of $F_\mathrm{middle}^{\mathrm{t}} = \exp \left(\log(F^{\mathrm{t}}(N=1)) - 1/2 *\left(-1/\xi_{0}^{\mathrm{th}}\right) \right)$. The decay rates $\xi_{0}$ and $N_{0}$ were obtained excluding the values for $N=1$ networks, and it was observed that this omission had a small effect on the fit of the data.

\section{Discussion}
\label{Discussion}

First we discuss the wave mitigation properties of the presented granular network geometry. By wave mitigation, we are referring to the ability of such a network to rapidly reduce the force amplitude of each pulse traveling through the system as compared to the incident pulse, i.e. reduce the relative impulse energy carried by each pulse. In our system, this is achieved by breaking the incident pulse up in to a series of smaller pulses spread both in space and time. The spatial mitigation is achieved through the branched geometry, and the temporal mitigation (i.e. series of decreasing amplitude pulses arriving at a given branch end) is achieved through the nonlinear dynamics properties of the system. The response is a result of the structured granular network geometry only, and does not require viscoelasticity or dissipation, such as plastic deformation, to achieve the force reduction.

\begin{figure}[!ht]
\includegraphics[scale=0.53]{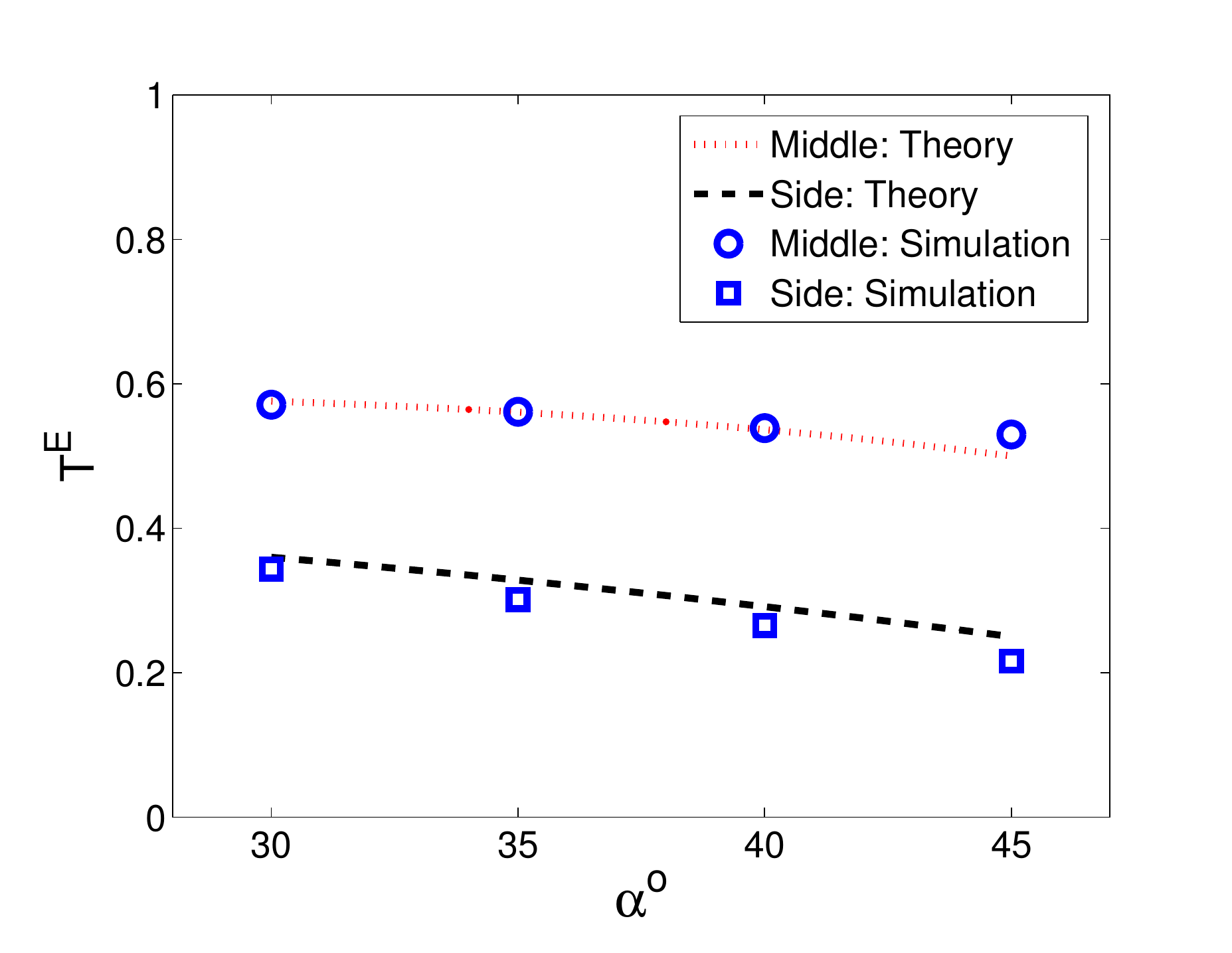}
\centering
\caption{Variations of the transmission coefficient $T^{\mathrm{E}}$ in terms of transferred acoustic energy through one stage of the granular chain network for the middle and side exit chains for both theoretical predictions (see Eq. (\ref{Eq_TE}) and (\ref{Eq_TEside})) and numerical simulations of the long branch (16 spheres per segment) networks with variable branching angles $\alpha$. The overall transmitted acoustic energy follows $E^{\mathrm{t}} = E^{\mathrm{i}} \left(T^{\mathrm{E}}\right)^N$.}\label{fig_TE}
\end{figure}

The main mechanism of wave mitigation in the granular chain networks is rather clear: for each additional degree added to the network, the leading pulse splits into two new pulses, so the acoustic energy of the leading pulse after $N$ branchings decays as
\begin{equation}
E^{\mathrm{t}}(N) = \left ( T^{\mathrm{E}} \right ) ^N E^{\mathrm{i}} = E^{\mathrm{i}} e^{-\frac{N}{N_0}},
\end{equation}
where we introduce here the transmission coefficient $T^{\mathrm{E}} < 1$ in terms of transferred acoustic energy through one stage of the granular chain network. Using the relationship between energy and force amplitude of solitary waves $E \sim F^{\nicefrac{5}{3}}$, the transmission coefficients derived in Section \ref{Quasi-Particle Theory} can be expressed in terms of energy, $\displaystyle T^{\mathrm{E}}_{\mathrm{K}} = \left ( T_{\mathrm{K}} \right )^{\nicefrac{5}{3}}$ where $\mathrm{K} \in \{\mathrm{S}, \mathrm{C}, \mathrm{M} \}$. A rough description of the granular chain network would be to assume that the acoustic energy divides into two equal parts at each new level, resulting in $\displaystyle T^{\mathrm{E}} = \nicefrac{1}{2}$ and $\displaystyle E^{\mathrm{t}}(N) = \frac{E^{\mathrm{i}}}{2^N}$. This would capture only qualitatively the overall behavior of the metamaterial designed here since wave recombinations as well as wave reflections can occur at branches and corners. As a result, the actual value of the effective transmission coefficient depends on the exit chain considered, and for the middle chain where the transmitted wave has the highest energy, one obtains
\begin{equation}
\displaystyle T^{\mathrm{E}}_{\mathrm{middle}} = \left( T_{\mathrm{S}} \sqrt{T_{\mathrm{C}} T_{\mathrm{M}}} \right )^{\nicefrac{5}{3}} = \frac{16 \cos^4 \alpha} {(1 + 2 \cos^2 \alpha)^3}
\label{Eq_TE}
\end{equation}
with $\displaystyle \nicefrac{1}{2} \leq T^{\mathrm{E}}_{\mathrm{middle}} < 0.58$ for the range of admissible branching angles $30^o \leq \alpha \leq 45^o$. This energy transmission coefficient is related to the rate $N_0$ of the exponential decay by the relation $N_0 = \frac{-5}{3 \log(T^{\mathrm{E}}_{\mathrm{middle}})}$.\footnote{We need to use here the relation between both types of transmission coefficient $\displaystyle T_{\mathrm{middle}}^{\mathrm{E}} = \left ( T_{\mathrm{middle}} \right )^{\nicefrac{5}{3}}$ and observe from Eq. (\ref{Eq_N0}) that $N_0 = -\frac{1}{\log(T_{\mathrm{middle}})}$ where $T_{\mathrm{middle}} = T_{\mathrm{S}} \sqrt{ T_{\mathrm{C}} T_{\mathrm{M}} }$ is the transmission coefficient through one stage of the network in terms of wave amplitude.}
Considering now the exit chains located on the very side of the network (see Fig. \ref{description}), one can show that the overall transmission coefficient that results from the propagation of the wave through the side branches of the network follows 
\begin{equation}
T^{\mathrm{E}}_{\mathrm{side}} = \left( T_{\mathrm{S}} T_{\mathrm{C}} \right)^{\nicefrac{5}{3}} = \frac{4 \cos^4 \alpha} {(1 + 2 \cos^2 \alpha)^2}
\label{Eq_TEside}
\end{equation}
and is comprised within $\displaystyle \nicefrac{1}{4} \leq T^{\mathrm{E}}_{\mathrm{side}} < 0.36$ for the range of admissible branching angles. Transmission coefficients for both middle and side branches are represented in Fig. \ref{fig_TE} as a function of the branching angle together with the results of simulations. In numerical simulations, $T^E_{\mathrm{middle}}$ and $T^E_{\mathrm{side}}$ were calculated from the fitted $N_{0}^{\mathrm{middle}}$ and $N_{0}^{\mathrm{side}}$ values obtained for long branch networks with $N=1$ though $N=6$ with $\alpha$ varying between $30^0$ and $45^0$, using the relation $N_{\mathrm{0}}= \frac{-5}{3 \log(T^{\mathrm{E}})} $. The agreement between the numerical simulations and the theoretical predictions of both $T^E_{\mathrm{middle}}$ and $T^E_{\mathrm{side}}$ is quite good.

\begin{figure}[!ht]
\includegraphics[scale=0.90]{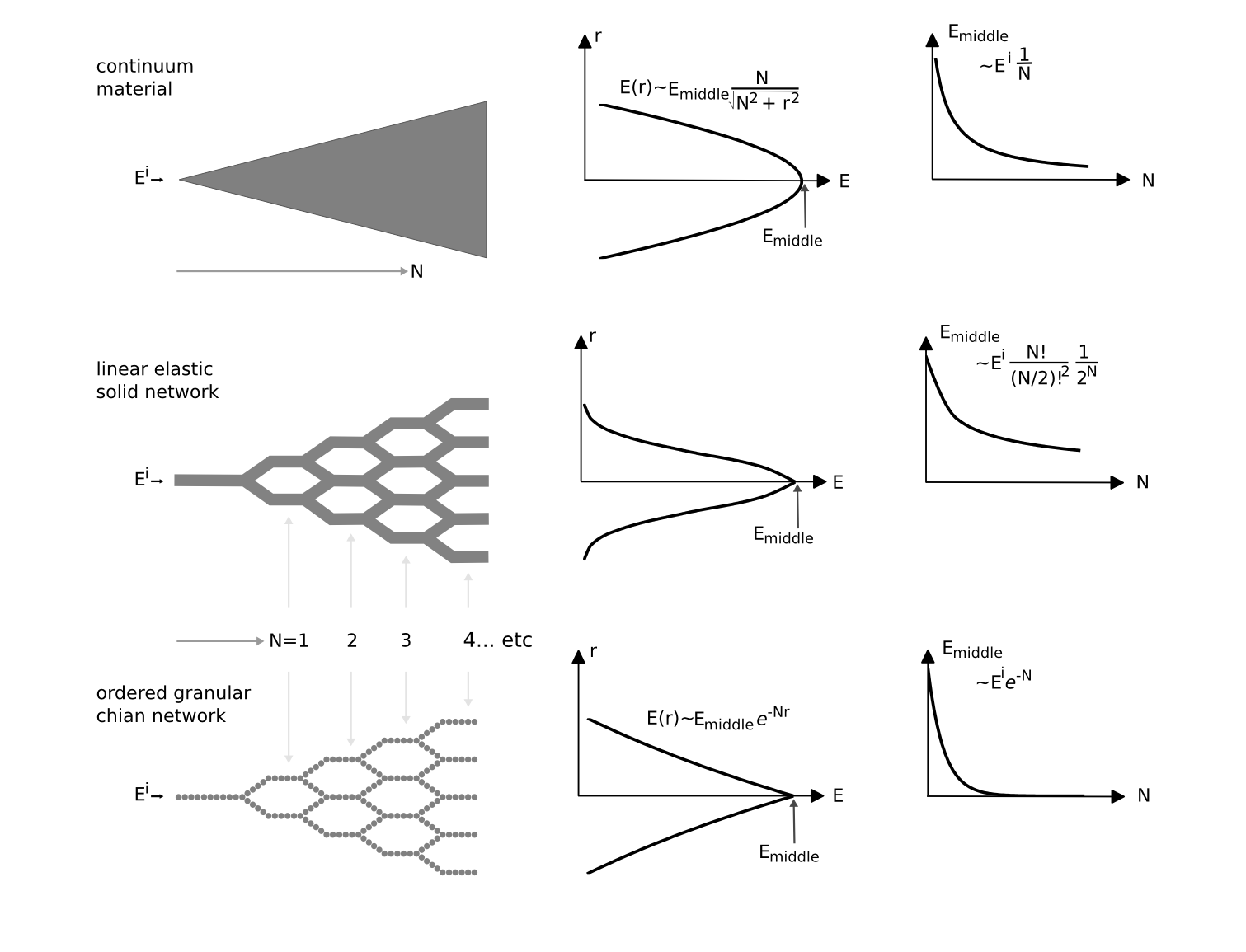}
\centering
\caption{(Top) Spatial distribution of the energy of a harmonic wave transmitted through a conventional 2D continuum material. (Middle) Spatial distribution of acoustic energy through an idealized solid network composed of a linear elastic material. (Bottom) Spatial distribution of acoustic energy in ordered granular networks (granular acoustic metamaterial).}\label{continuum}
\end{figure}

The remarkable ability of granular chain networks to efficiently mitigate the acoustic energy of the largest wave in the system emerges from their branched structure. Due to the underlying branched structure, the spatial distribution of acoustic energy is also fundamentally different than the wave front energy distribution through a conventional continuum material (see Fig. \ref{continuum}). In a non-dissipative 2D elastic medium, the elastic energy density of a harmonic wave is shown to decay as $1/N$ with the distance $N$ from a localized excitation, invoking energy conservation. More generally, a power law decay of the elastic energy is expected for either linear or nonlinear acoustic media. Similarly, the wave spatial structure at a distance $N$ follows a power law behavior, as shown in Fig. \ref{continuum} \, (Top). Conversely, the granular chain network exhibits a drastically faster decay of the energy carried in the leading pulse: exponential decay of both the central leading pulse with propagation distance and the spatial repartition of the transmitted wave with the distance to the central axis (Fig. \ref{continuum} \, (Bottom)). The theoretical predictions also indicate that the exponential decay occurs over relatively short length scales. Specifically, the values of the decay length scales $N_{0}$ and $\xi_{0}$ suggest that the $N=3$ network is already a highly effective acoustic wave mitigating structure. To better understand the importance of the nonlinear dynamic response in the wave mitigating properties of our system, we also compare our granular network with its static response and, similarly, with the dynamic response of a linear elastic solid network (Fig. \ref{continuum}).

If we consider a static load applied to the input branch, because of the symmetry of our system, the compressive forces in each segment can be calculated assuming an even force splitting at all nodes. The static forces carried by each branch will be largest in the center of the system and decrease with increasing N, and decrease with increasing perpendicular distance r. For the studied network geometry, we can write a general expression for the maximum member force as well as for the smallest (side or r=0.5) force. 
\begin{equation}
\left\{
\begin{array}{l c c}
\displaystyle F_{\mathrm{max}}^{\mathrm{static}}(N) & = & F^{\mathrm{i}} \frac{N!}{\frac{N}{2}!^2} \frac{1}{2^{N}}  \quad \mathrm{for \, } N \mathrm{\, even \, networks}
\vspace{10pt}
\\
\displaystyle F_{\mathrm{max}}^{\mathrm{static}}(N) & = & F^{\mathrm{i}}  \frac{N!}{\frac{N-1}{2}!\frac{N+1}{2}!} \frac{1}{2^{N}}  \quad \mathrm{for \, } N \mathrm{\, odd \, networks}
\vspace{10pt}
\\
\displaystyle F_{\mathrm{side}}^{\mathrm{static}}(N,r=0.5) & = & F^{\mathrm{i}} \frac{1}{2^{N}}  \quad \mathrm{~~~~~~for~all~\, } N~~~~~~~~~~~~~~~
\end{array}
\right .
\label{static}
\end{equation}
While the side branch force decays exponentially, the middle (largest) force decays with a rate between linear and exponential, as $N$ increases. The static response of our granular network structures is the same as a branched network composed of any homogeneous solid media. However, the dynamic response of a homogeneous solid network does not differ from its static response in terms of relative force transmission, assuming a linear elastic, non-dispersive media, and neglecting complexities such as edge effects. This comparison clearly shows that the exponential decay in the largest leading pulse in our granular network is a result of the structured branching geometry in combination with the nonlinear dynamic response of our system (Fig. \ref{continuum}).

The analytical procedure presented here is general, and could be used to describe more complex systems with any combination of material properties and sizes in adjacent branches. In a more general system, the transmission coefficients, and therefore overall effective response, would also depend on the particle sizes and material properties. The presented network structure is not intended to be the most efficient granular network. On the contrary, we study a simplified (i.e. symmetric and homogeneous) network, in order to provide and validate a method for designing more complex granular network structures. For example, chains of different materials could be introduced within the network to tune the wave velocity and prevent pulse recombinations and to increase the ratio of reflected to transmitted wave amplitudes. Previous studies have shown that adding variable particle sizes or material properties within 1D bead chains can be used to significantly reduce the incident pulse \cite{Sen:2008}. Additionally, a honeycomb structure could be designed to produce mitigating behavior when the system is excited from either the left or right side. An additional benefit of the granular networks is the low effective density compared to a conventional bulk material. Here, we study the propagation of a single incident pules. However, one of the attractive wave mitigating features of granular chains is that they naturally break up large or long duration excitations into a series of smaller pulses spread in time, by opening gaps between particles (see \cite{Sen:2008}). For example, the entrance chain of our network could be used to break up a shock like excitations, subsequently each pulse resulting from the impact would be further reduced as it propagates through the network. Such a system could also be impacted with much larger forces. However, one of the benefits of the proposed granular structure is its reversible deformation, a feature that would be lost for high force impacts, but with the trade off of increased force mitigation resulting from the plastic deformation. Such techniques could be incorporated into our network structure to further improve the force mitigating properties.

Beyond these interests for engineering applications, our model system also provides new insights into wave propagation in disordered granular media. The dynamic behavior of granular materials is not well understood, yet it plays an important role in many areas of industry and research ranging from agriculture and construction to modeling earthquakes and avalanches. In tightly packed granular media, several previous studies showed that the dynamic force transmission occurs along force chains \cite{Bardenhagen:1998,Roessig:2002}, following preferred loading paths within the contact network. \cite{Owens:2011} suggest that these granular chain network comes from the static force chains widely observed in granular packings at rest \cite{Liu:1995}. Our system contains all the relevant ingredients to capture the physics of wave propagation though force chain networks: wave splitting, bending, and recombination. As a result, we expect the ordered granular network presented here to qualitatively capture the behavior of disordered granular media. In terms of effective acoustic response, the exponential mitigation of the wave amplitude identified in this study for ordered granular packings is also observed in disordered packings \cite{Clark:2012,Owens:2011}. In addition, the spatio-temporal structure of the wave transmitted through ordered and disordered granular media shares qualitatively many common features: after a dominant leading pulse that has been the main focus of our study, we also observe a train of smaller solitary waves resulting from wave propagation through alternative and less direct paths (see \cite{Jia:2004} for similar observations in random packings). Interestingly, due to pulse recombinations, it is not impossible to observe that the largest transmitted pulse is actually not the leading one for some special network configurations, as also reported for random networks. Let us note that dissipation present in real systems will affect the actual rate of mitigation. Comparing our dissipative experiments with the conservative numerical simulations, we observed that the rate of mitigation due to wave splitting and dissipation are of the same order for the short branch networks. This ingredient should also be included in a realistic model of wave propagation in natural granular materials. Finally, since our granular chain system was perfectly ordered, we could investigate its properties analytically, and relate its property at the microstructure scale (branching angle and particle geometry) with its macroscopic acoustic behavior ($N_0$ and $\xi_0$). In our experiments, the particle alignment was crucial in order to validate the predicted properties for the specific network geometry chosen. However, future studies involving branched systems that incorporate random orientations and chain lengths would help to improve our understanding of the acoustic transmission through disordered force chain networks and identify the main differences with ordered networks. We could, for example, calculate an effective disordered network geometry, and subsequently use the effective geometry to predict the disordered system response. Or, perhaps an effective chain length and branching angle can be established, based on the level of disorder, which could then be used to describe an effective acoustic behavior of disordered networks using a similar theoretical approach as for the ordered network. Such future studies would help elucidate the role of disorder in the wave mitigation capabilities of granular media.

\section{Conclusions}
We studied the transmission of elastic waves through ordered granular chain networks with variable branch angles. We identified three main physical mechanisms involved during this process: wave splitting, wave bending through corners, and wave recombination at merging junctions. Solitary wave splitting was previously observed \cite{Daraio:2010,Ngo:2012}, however, we present the first observation and theoretical description of two identical solitary waves combining. The quasi-particle theory was used to describe these mechanisms, and derive the effective acoustic properties of the granular acoustic metamaterial from the geometrical properties of the network and the bead properties. Numerical simulations for long branch networks up to branching level $N=6$ are in good quantitative agreement with our theoretical approach for both the exponential decay of the central leading pulse with propagation distance and the spatial repartition of the transmitted wave with the distance to the central axis. Experiments and numerical simulations of a short branch system for $N$ up to 4 are also in good qualitative agreement with the predicted exponential decay of the leading central pulse amplitude and the spatial repartition of the leading pulses along $r$. However, the short branch system does not always provide a long enough propagation distance for leading and trailing pulses to separate, and pulse overlapping results in an increased wave amplitude compared to the quasi-particle predictions which only account for the leading pulses. Additionally, the effects of dissipation present in experiments reduce the leading pulse amplitude faster than in our theoretical model. Overall, the exponential decay of leading pulse amplitudes through the ordered granular network makes these systems ideally suited for wave mitigation applications. Perhaps most importantly, the good agreement between experiments and both numerical simulations and theoretical predictions suggests that more complex designs with controllable wave propagation pathways should also be experimentally feasible. Furthermore, the exponential decay of wave amplitude with propagation distance in our ordered granular network compares well with the wave propagation features along force chains in disordered granular media \cite{Clark:2012,Owens:2011}. This ordered version of natural granular materials offers the possibility to understand in all its details the acoustic behavior of a granular medium and (i) relate microstructural characteristics with effective acoustic properties in this type of materials and (ii) choose adequately the material microstructure to achieve desired acoustic response like a high mitigation rate.

\section*{Acknowledgements}
This work was supported in part by the Department of Energy Office of Science Graduate Fellowship Program (DOE SCGF), made possible in part by the American Recovery and Reinvestment Act of 2009, administrated by ORISE-ORAU under contract DE-AC05-06OR23100. Laurent Ponson gratefully ackowledges the support of the European Union through the Marie Curie integration grant "ToughBridge".


\section*{References}


\begin{thebibliography}{99}

\bibitem{Pendry}
Pendry, J., 2000. Negative refraction makes a perfect lens. Phys. Rev. Lett.
  85, 3966--3969.
  
\bibitem{Shelby}
Shelby, R.~A., Smith, D.~R., Schultz, S., 2001. Experimental verification of a
  negative index of refraction. Science 292, 77--79.

\bibitem{Fang}
Fang, N., Xi, D., Xu, J., Ambati, M., Srituravanich, W., Sun, C., Zhang, X.,
  2006. Ultrasonic metamaterials with negative modulus. Nature Mater. 5,
  452--456.
 
  
\bibitem{Brun}
Brun, M., Guenneau, S., Movchan, A., Bigoni, D., 2010. Dynamics of structural
  interfaces: Filtering and focussing effects for elastic waves. Journal of the
  Mechanics and Physics of Solids 58~(9), 1212 -- 1224.
   
\bibitem{Huang}
Huang, H., Sun, C., 2011. Theoretical investigation of the behavior of an
  acoustic metamaterial with extreme young's modulus. Journal of the Mechanics
  and Physics of Solids 59~(10), 2070 -- 2081.
   
\bibitem{Huang2}
Huang, H.~H., Sun, C.~T., 2009. Wave attenuation mechanism in an acoustic
  metamaterial with negative effective mass density. New J. Phys. 11, 013003.
  
\bibitem{Mei}
Mei, J., Ma, G., Yang, M., Yang, Z., Wen, W., Sheng, P., 2012. Dark acoustic
  metamaterials as super absorbers for low-frequency sound. Nature Com. 3, 756.

\bibitem{Popa}
Popa, B.~I., Cummer, S.~A., 2009. Design and charcterization o broadband
  acoustic composite metamaterials. Phys. Rev. E 80, 174303.

\bibitem{Nesterenko:1983}
Nesterenko, V., 1983. Propagation of nonlinear compression pulses in granular
  media. J. Appl. Mech. Tech. Phys. 24~(5), 733--743.

\bibitem{FNesterenko:2001}
Nesterenko, V., Jan 2001. Dynamics of heterogeneous materials. Springer-Verlag
  New York, Inc.
  
\bibitem{Coste:1997}
Coste, C., Falcon, E., Fauve, S., 1997. Solitary waves in a chain of beads
  under hertz contact. Phys. Rev. E 56~(5), 6104--6117.

\bibitem{Daraio:2005}
Daraio, C., Nesterenko, V., Herbold, E., Jin, S., 2005. Strongly nonlinear
  waves in a chain of teflon beads. Phys. Rev. E 72~(1), 016603.  
  
\bibitem{Sen:2008}
Sen, S., Hong, J., Bang, J., Avalos, E., Doney, R., 2008. Solitary waves in the
  granular chain. Phys. Rep. 462~(2), 21 -- 66.
  
\bibitem{Job:2007}
Job, S., Melo, F., Sokolow, A., Sen, S., 2007. Solitary wave trains in granular
  chains: experiments, theory and simulations. Granular Matter 10, 13--20.
  
\bibitem{Daraio:2010}
Daraio, C., Ngo, D., Nesterenko, V.~F., Fraternali, F., Sep 2010. Highly
  nonlinear pulse splitting and recombination in a two-dimensional granular
  network. Phys. Rev. E 82, 036603.

\bibitem{Makse}
Makse, H.~A., Gland, N., Johnson, D.~L., Schwartz, M., 1999. Why effective
  medium theory fails in granular materials, 5070--5073.

\bibitem{Liu:1995}
Liu, C.~h., Nagel, S.~R., Schecter, D.~A., Coppersmith, S.~N., Majumdar, S.,
  Narayan, O., Witten, T.~A., 1995. Force fluctuations in bead packs. Science
  269~(5223), 513--515.
  
\bibitem{Bardenhagen:1998}
Bardenhagen, S.~G., Brackbill, J.~U., 1998. Dynamic stress bridging in granular
  material. Journal of Applied Physics 83~(11), 5732 --5740.

\bibitem{Clark:2012}
Clark, A. H., Kondic, L., and Behringer, R. P., 2012. Particle Scale Dynamics in Granular Impact. Phys. Rev. Lett. \textbf{109} 238302.

\bibitem{Roessig:2002}
Roessig, K., Foster, J., Bardenhagen, S., 2002. Dynamic stress chain formation
  in a two-dimensional particle bed. Experimental Mechanics 42, 329--337,
  10.1007/BF02410990.
  
  \bibitem{Owens:2011}
Owens, E.~T., Daniels, K.~E., 2011. Sound propagation and force chains in
  granular materials. Eur. Phys. Lett. 94, 54005.
  
  \bibitem{Shukla}
Shukla, A., Zhu, C., Sadd, M., 1988. Angular dependence of dynamic load
  transfer due to explosive loading in granular aggregate chains. The Journal
  of Strain Analysis for Engineering Design 23~(3), 121--127.
  
  \bibitem{Ngo:2012}
Ngo, D., Fraternali, F., Daraio, C., Mar 2012. Highly nonlinear solitary wave
  propagation in y-shaped granular crystals with variable branch angles. Phys.
  Rev. E 85, 036602.
  
  \bibitem{Qiong:2013}
Qiong, C., Xian-Qing, Y., Xin-Yin, Z., Zhen-Hui, W., Yue-Min, Z., 2013. Binary collision approximation for solitary waves in a Y-shaped granular chain. \textit{Chinese Physics B} \textbf{22}(1), 014501.

\bibitem{Job:2005}
Job, S., Melo, F., Sokolow, A., Sen, S., May 2005. How hertzian solitary waves
  interact with boundaries in a 1d granular medium. Phys. Rev. Lett. 94,
  178002.
  
\bibitem{Cundall:1979}
Cundall, P.~A., Strack, O. D.~L., 1979. A discrete numerical model for granular
  assemblies. Geotechnique 29, 47--65.

\bibitem{Johnson:1987}
Johnson, K., Jan 1987. Contact mechanics. The Press Syndicate of the University
  of Cambridge, The Pitt Building.

\bibitem{CarreteroGonzalez:2009}
Carretero-Gonz{\'a}lez, R., Khatri, D., Porter, M., Kevrekidis, P., Daraio, C.,
  Jan 2009. Dissipative solitary waves in granular crystals. Phys. Rev. Lett.
  102~(2), 024102.

\bibitem{Herbold:2010}
Herbold, E.~B., Nesterenko, V.~F., 2010. The role of dissipation on wave shape
  and attenuation in granular chains. Physics Procedia 3~(1), 465 -- 471.

\bibitem{Rosas:2007}
Rosas, A., Romero, A.~H., Nesterenko, V.~F., Lindenberg, K., Apr 2007.
  Observation of two-wave structure in strongly nonlinear dissipative granular
  chains. Phys. Rev. Lett. 98, 164301.

\bibitem{Jia:2004}
Jia, X., 2004. Codalike multiple scattering of elastic waves in dense granular
  media. Phys. Rev. Lett. 93, 154303.

\bibitem{Jia2}
Jia, X., Caroli, C., Velicky, B., 1999. Ultrasound propagation in externally
  stressed granular media. Phys. Rev. Lett. 82, 1863--1866.




 



\end{thebibliography}
\end{document}